# Thickness-Dependent Polaron Crossover in Tellurene

Kunyan Zhang[1,2]†, Chuliang Fu[3]†, Shelly Kelly[4], Liangbo Liang[5], Seoung-Hun Kang[6], Jing Jiang[7], Ruifang Zhang[7], Yixiu Wang[7], Gang Wan[8], Phum Siriviboon[9], Mina Yoon[6], Peide Ye[10], Wenzhuo Wu[7], Mingda Li[3]*, Shengxi Huang[11]*

[1]Molecular Biophysics and Integrated Bioimaging Division, Lawrence Berkeley National Laboratory, Berkeley, CA 94720, USA
[2]Department of Chemistry, University of California, Berkeley, CA 94720, USA
[3]Department of Nuclear Science and Engineering, Massachusetts Institute of Technology, Cambridge, MA 02139, USA
[4]X-ray Science Division, Argonne National Laboratory, Lemont, IL 60439, USA
[5]Center for Nanophase Materials Sciences, Oak Ridge National Laboratory, Oak Ridge, TN 37831, USA
[6]Materials Science and Technology Division, Oak Ridge National Laboratory, Oak Ridge, TN 37831, USA
[7]School of Industrial Engineering, Purdue University, West Lafayette, IN, 47907, USA
[8]Department of Mechanical Engineering, Stanford University, Stanford, California 94305, USA
[9]Department of Physics, Massachusetts Institute of Technology, Cambridge, MA 02139, USA
[10]Elmore Family School of Electrical and Computer Engineering, Purdue University, West Lafayette, IN 47907, USA
[11]Department of Electrical and Computer Engineering, Rice University, Houston, TX 77005, USA

*Corresponding author: Shengxi Huang (shengxi.huang@rice.edu), Mingda Li (mingda@mit.edu).
†These authors contributed equally to this work.

## Abstract
Polarons, quasiparticles arising from electron-phonon coupling, are crucial in understanding material properties such as high-temperature superconductivity and colossal magnetoresistance. However, scarce studies have been performed to investigate the formation of polarons in low-dimensional materials with phonon polarity and electronic structure transitions. In this work, we studied polarons of tellurene that are composed of chiral chains of tellurium atoms. The frequency and linewidth of the $A_1$ phonon, which becomes increasingly polar for thinner tellurene, exhibit an abrupt change when the thickness of tellurene is below 10 nm. Meanwhile, the field effect mobility of tellurene drops rapidly as the thickness is smaller than 10 nm. These phonon and transport signatures, combined with the calculated phonon polarity and band structure, suggest a crossover from large polarons for bulk tellurium to small polarons for few-layer tellurene. Effective field theory considers the phonon renormalization in the strong coupling (small polaron) regime, and semi-quantitatively reproduces the observed phonon hardening and broadening effects in few-layer tellurene. This polaron crossover stems from the quasi-1D nature of tellurene where modulation of the interchain distance reduces the dielectric screening and promotes electron-phonon coupling. Our work provides valuable insights into the influence of polarons on phononic, electronic, and structural properties in low-dimensional materials.

# Introduction

Polaron is a quasiparticle formed by the coupling between charge carriers and lattice vibrations (phonons). The interaction between charge carriers and phonons induces a polarization cloud that follows the charge carriers, which can lead to higher effective mass, lower mobility, and hopping-like conductivity, distinct from nearly-free electrons (*1-4*). Because of its unique features, polaron plays a crucial role in various physical properties, such as electronic transport, magnetoresistance, ferroelectricity, and thermoelectricity (*5, 6*). For example, the strong Jahn-Teller effect in $La_{1-x}Ca_xMnO_{3+y}$ can lead to the formation of polarons which is associated with the high-temperature superconductivity (*7-9*). Depending on the coupling strength between electrons and phonons, polarons can be categorized into small polarons and large polarons based on the spatial distribution (*5*). For small polarons with small spatial extension, the electron-phonon coupling is strong compared to the kinetic energy of electrons. In this strong coupling regime, the electron becomes localized within a regime of a size comparable to the unit cell, forming small polarons (*3, 10*). In contrast, large polarons with large polaron radii are formed in the weak coupling regime where the electron can delocalize over a larger region in the lattice (*11*). In solid-state materials, the crossover from large polarons to small polarons (weak coupling to strong coupling) is often associated with metal-to-semiconductor/insulator transitions, with scattered reports insofar (*12, 13*). In $La_{1-x}Ca_xMnO_3$, the transition from large polaron at low temperature to small polaron at high temperature can be revealed by the characteristic Raman scattering response at metal-to-semiconductor transition (*13*). This polaron crossover has also been quantitatively measured by the Jahn-Teller polaronic distortion using extended X-ray absorption fine structure (EXAFS) analysis where the Mo-O bond length showed noticeable changes (*12*).

Tellurene serves as an ideal platform to study polarons because of the relatively large polaron coupling constant (*14*). Tellurene is composed of helical chains of tellurium atoms and hosts unique properties which make it promising for applications such as broadband photodetectors (*15, 16*), high-mobility field effect transistors (FET) (*17*), and topological phase change transistors (*18*). The carrier mobility of tellurene FET shows a dependence on the film thickness, consistent with the increase of band gap from bulk tellurium to few-layer tellurene (*17*). The dependence of electronic structure on thickness also leads to different topological states at varied temperatures. In topological phase change transistors based on tellurene, thick tellurene of 32 nm exhibits negative magnetoresistance below 100 K, a signature of topological states, while thin tellurene of 12 nm shows only positive magnetoresistance (*18*). These transport behaviors modulated by tellurene thickness suggest that the polaronic properties of tellurene can significantly depend on its thickness.

In this work, we investigate the polarons of tellurene and the crossover from large polaron to small polaron as tellurene thickness reduces. In thinner tellurene, the longitudinal $A_1$ phonon becomes polar as opposed to non-polar in bulk tellurium, as shown by density-functional theory (DFT) calculations. This distinct increase in phonon polarity, a requirement for forming small polarons, strongly indicates the presence of small polaron/large polaron crossover dependent on tellurene thickness. The transition from large polarons in bulk tellurium to small polarons in few-layer tellurene is also exhibited in our measured $A_1$ phonon, which shows a blueshift by 10 cm$^{-1}$ and a broadening by 5 cm$^{-1}$ for tellurene thickness below 10 nm. To understand the influence of polarons on phonon properties, we developed a theoretical model to explain the phonon hardening and broadening effect due to polaron formation. Specifically, the small polaron contributes to strong renormalization of phonon frequency and linewidth, which is absent in the large polaron regime. This polaron crossover as a function of tellurene thickness can be understood by the structural transition of tellurene as its thickness reduces. EXAFS spectroscopy reveals that the distance between neighboring Te chains changes drastically for smaller thicknesses, which is in agreement with the formation of small polarons due to reduced dielectric screening and enhanced electron-phonon coupling. Our unified experimental, computational, and theoretical study provides a comprehensive picture of the thickness-dependent polaron formations in few-layer tellurene. The understanding of polaron formation and its implication for electronic structure and phonon response is fundamental for tailoring the electronic properties of tellurene-based devices.

## Results
### Thickness dependence of Raman signatures in few-layer tellurene

Tellurene is the two-dimensional (2D) form of bulk tellurium, which is composed of chiral chains of tellurium atoms in a hexagonal lattice as shown in Figs. 1A and 1C. Each tellurium atom is covalently bonded to the two nearest atoms within the same chain, while the neighboring chains interact through weak van der Waals force (*19*). The few-layer tellurene synthesized by the hydrothermal method (*17*) has a typical thickness of a few nanometers and a lateral dimension in micrometers as measured by atomic force microscopy (Fig. 1B and fig. S1). As shown in Fig. 1D, bulk tellurium exhibits the $A_1$ mode at 121 cm$^{-1}$ and the $E_2$ mode at 141 cm$^{-1}$. These phonon modes correspond to the chain expansion in the basal plane and asymmetric stretching along the c-axis, respectively. The polarization-dependent measurement in fig. S2 reveals the phonon symmetry, which is consistent with prior reports on α-phase tellurium (*20*). As the thickness of tellurene reduces, the $A_1$ mode exhibits a blue shift of more than 10 cm$^{-1}$ accompanied by a broadening of the linewidth (Fig. 1D). Figures 1E-G summarize the frequency, linewidth, and asymmetry of the $A_1$ phonon as a function of tellurene thickness by fitting with a Breit-Wigner-Fano (BWF) function (Supplementary Note 1). For tellurene with a thickness from 20 nm to 5 nm, the frequency of the $A_1$ phonon increases from 121 cm$^{-1}$ to 133 cm$^{-1}$ (Fig. 1E). The overall increase of $A_1$ phonon frequency can be partly understood by the increased intrachain bond strength for thinner layers, which contributes to a more effective restoring force for the chain expansion of the $A_1$ mode. However, this cannot explain the drastic transition of phonon frequency at the thickness of $10 \pm 2$ nm (Fig. 1E). At the same critical thickness, the linewidth of the $A_1$ mode also shows considerable increases approximately from 2.5 to 5 cm$^{-1}$ (Fig. 1F), suggesting a change in the phonon lifetime. Additionally, the asymmetry parameter $1/q_s$ becomes proportional to the thickness below 10 nm, with a maximum $1/q_s = 0.28$ for a 4 nm tellurene as shown in Fig. 1G. The asymmetric lineshape of a phonon mode is a signature of the Fano resonance effect where the scattering of an electronic continuum is interfering with the discrete phonon mode (*21, 22*). The observed Fano resonance in the $A_1$ phonon indicates that tellurene shows strong electron-phonon coupling below 10 nm. These characteristic phonon features can be understood by the formation of small polarons in tellurene because of reduced dielectric screening below the critical thickness. As the thickness of tellurene reduces, the less dielectric screening leads to enhanced electron-phonon coupling, which is evidenced by the asymmetry of the $A_1$ mode (Fig. 1G). Moreover, the formation of small polarons raises the electrical resistance due to strong electron-phonon interaction to localize the electrons. This is consistent with the changes in field-effect mobility of tellurene as shown in Fig. 1H, which experiences an immediate decrease around $10 \pm 2$ nm from 600 cm$^2$ V$^{-1}$ s$^{-1}$ down to 50 cm$^2$ V$^{-1}$ s$^{-1}$ (*17*).

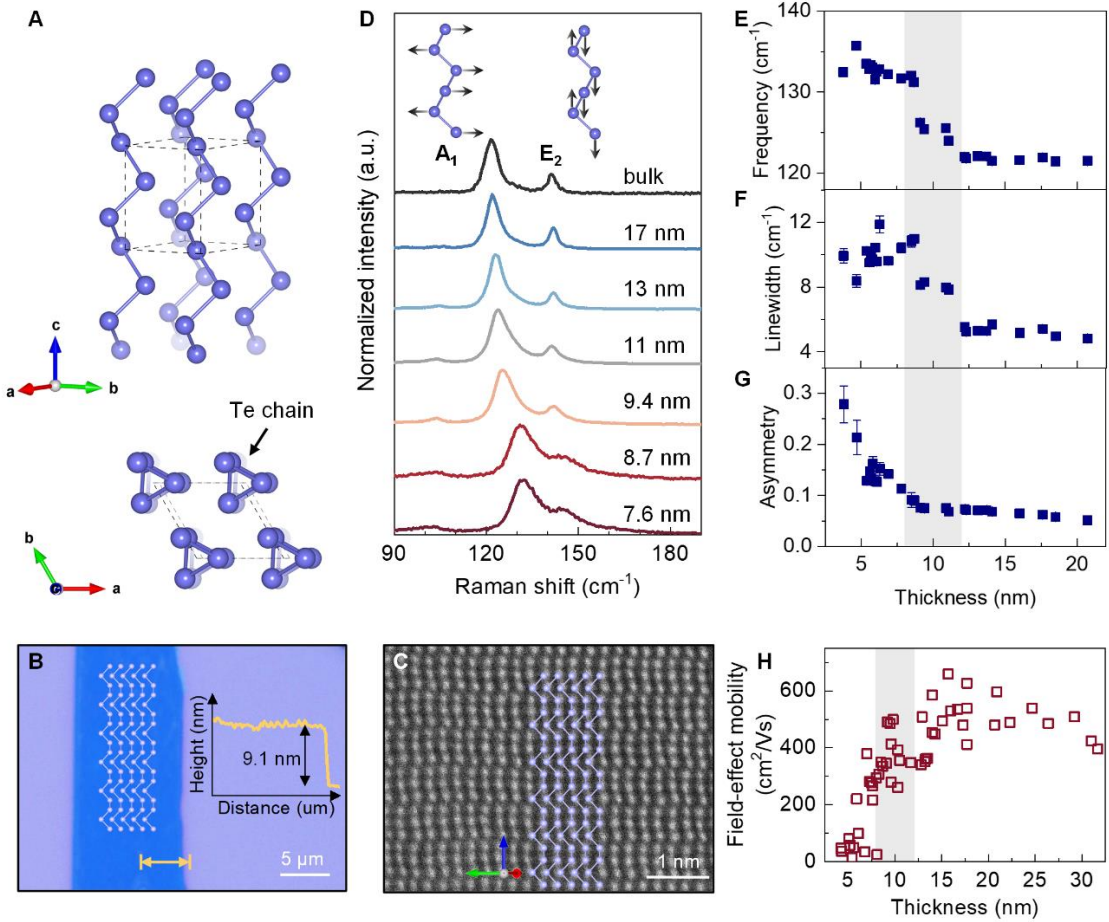

**Fig. 1. Thickness-dependent structural, phononic, and electronic properties of tellurene.** (**A**) The crystal structure of tellurene. (**B**) Optical image of tellurene with a thickness of 9.1 nm. Inset: depth profile across the arrow. (**C**) Scanning transmission electron microscopy (STEM) images of tellurene. (**D**) Raman spectra for tellurene with different thicknesses. Inset: lattice vibration of the $A_1$ and $E_2$ phonon modes. (**E-G**) The $A_1$ phonon frequency, linewidth, and asymmetry as a function of thickness. (**H**) The field-effect mobility of tellurene as a function of thickness.

## Thickness-dependent phonon polarity and bandgap: a first-principles study

To understand the change of the $A_1$ phonon, we performed first-principles calculations of the phonon properties and electronic structure. The overall blue shift of the phonon modes in few-layer tellurene as compared to bulk tellurium can be reproduced by calculations (Fig. 2A and fig. S3). According to a prior experimental study (*14*), among the four optical phonon modes at the Gamma point in bulk Te, only the $A_1$ phonon is non-polar and does not change the electric dipole moment (or polarization) of the system. This is confirmed by our calculations where the change of dipole moment caused by the $A_1$ phonon is zero in bulk tellurium (Fig. 2B). However, our calculations find out that the $A_1$ phonon becomes increasingly polar with decreasing thickness, as shown in Fig. 2B, where the change of the dipole moment caused by the $A_1$ phonon becomes drastically larger in few-layer tellurene as compared to bulk tellurium. As the $A_1$ mode vibrates in the *ab* plane, which is perpendicular to the chiral chain direction along the *c* axis (Fig. 1A), the change of the dipole moment is also perpendicular to the chain. In bulk tellurium, the contributions of Te atoms to the change of dipole moment by the $A_1$ phonon perfectly cancel out one another as illustrated in Figs. 2E-F (vibrations of different Te atoms are symmetrical to one another). With a decreased number of layers, the exterior Te atoms become increasingly non-equivalent to the interior Te atoms (*23*); therefore, their contributions to the change of dipole moment can no longer be canceled out by the interior Te atoms. This can be seen from the non-symmetrical vibrations of Te atoms in 2L tellurene (Figs. 2C-D). Our calculations suggest a non-polar to polar transition that occurs at a certain thickness for the $A_1$ mode, in contrast to the other three optical phonons ($E_1$, $E_2$, and $A_2$) that are always polar irrespective of the thickness (Fig. 2B). Such difference could contribute to the unique thickness dependence of the $A_1$ mode, especially the dramatic change of its frequency and linewidth at a certain sample thickness. In addition, the calculated band structure shows the transition from semimetal to semiconductor, where the calculated bandgap of 0.23 eV of bulk tellurium increases to 1.363 eV in bilayer tellurene (Figs. 2G-I and fig. S4), leading to a thickness-dependent optical absorption (fig. S5). The corroborated evolutions of phonon property and electronic structure point to the formation of small polarons in few-layer tellurene, where the localized small polarons cause the renormalization of phonon frequency and linewidth as compared to large polarons in bulk tellurium.

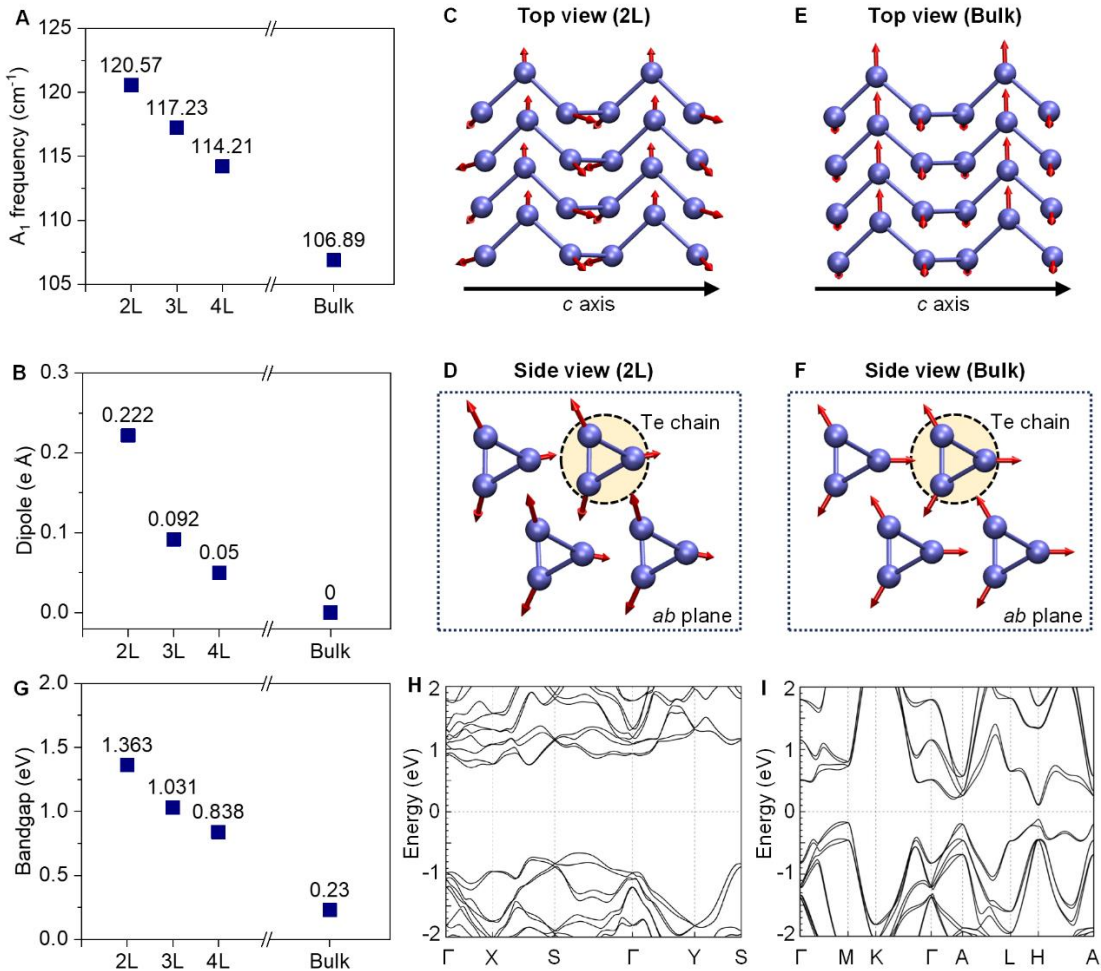

**Fig. 2.** Calculated phonon polarity and band structure for few-layer tellurene and bulk tellurium. (**A**) The calculated $A_1$ phonon frequency. (**B**) The calculated change of the dipole moment by the $A_1$ mode as a function of thickness. (**C-F**) Top view and side view with respect to the experiment geometry showing the calculated lattice vibrations of the $A_1$ mode in 2L tellurene and bulk tellurium. The red arrows represent the atomic vibrations. (**G**) The calculated bandgap of tellurene as a function of thickness. Calculated band structure of (**H**) 2L tellurene and (**I**) bulk tellurium.

**Polaron theory for phonon renormalization**

To describe the effect of polaron formation on phonon dispersion and linewidth, we start with a Hamiltonian consisting of three parts (*24*)

$$H = -t \sum_{\langle ij \rangle} c_i^\dagger c_j + \sum_{\mathbf{q}} \omega_{\mathbf{q}} a_{\mathbf{q}}^\dagger a_{\mathbf{q}} + \sum_{j\mathbf{q}} g_{\mathbf{q}} c_j^\dagger c_j (a_{\mathbf{q}} + a_{-\mathbf{q}}^\dagger) e^{i\mathbf{q}\cdot\mathbf{R_j}}. \tag{1}$$

The first two terms represent the noninteracting electrons and noninteracting phonons where $t$ is the electron hopping, $\omega_{\mathbf{q}}$ is the phonon energy at wavevector $\mathbf{q}$ and we have normalized $\hbar$ to be 1. Here we use $c_i^\dagger(c_i)$, $a_{\mathbf{q}}^\dagger(a_{\mathbf{q}})$ to denote the creation (annihilation) operator of electron at site $i$ and phonon at momentum $\mathbf{q}$, respectively. The third term denotes the electron-phonon coupling with $g_{\mathbf{q}}$ representing the coupling strength. This Hamiltonian can describe the behavior of polarons, as detailed in ref. (*24*). Depending on the strength of electron-phonon coupling, polarons can be classified into small polaron (with strong electron-phonon coupling) and large polaron (with weak electron-phonon coupling) as illustrated in Fig. 3A. In our case, the thinner Te sample with polar phonon and strong electron-phonon coupling prefers to form small polarons while the thicker Te sample forms large polarons. For the large polaron within the weak coupling regime, the change of the phonon frequency and linewidth is derived at the order of magnitude proportional to $|g_{\mathbf{q}}|^2$. Since $g_{\mathbf{q}}$ is relatively small in the weak coupling regime for large polarons, the phonon frequency blueshift should be small, and the phonon linewidth is narrower for thicker tellurene. Therefore, theoretically, large polarons do not significantly impact phonon properties, whereas small polarons can influence both phonon frequency and linewidth (Details in Supplementary Note 2).

For the small polaron within the strong coupling regime, the Hamiltonian can be rewritten using the Lang-Firsov canonical transformation (*24*) where the electron operator is transformed into the polaron operator

$$H = -\sum_j E_B c_j^+ c_j + \sum_{\mathbf{q}} \omega_{\mathbf{q}} a_{\mathbf{q}}^+ a_{\mathbf{q}} - t \sum_{\langle ij \rangle} exp\left(-\sum_{\mathbf{q}} T_{\mathbf{q}ij} p_{\mathbf{q}} e^{-i\mathbf{q}\cdot\mathbf{R}_i}\right) c_i^+ c_j, \tag{2}$$

where $T_{\mathbf{q}ij} = \frac{g_{\mathbf{q}}^*}{\omega_{\mathbf{q}}} i \sqrt{\frac{2}{m\omega_{\mathbf{q}}}}(1 - e^{i\mathbf{q}\cdot(\mathbf{R}_i - \mathbf{R}_j)})$ represents the lattice distortion due to the polaron and $p_{\mathbf{q}} = i\sqrt{\frac{m\omega_{\mathbf{q}}}{2}}(a_{\mathbf{q}}^+ - a_{-\mathbf{q}})$ is the momentum operator of the phonon, $E_B = \sum_{\mathbf{q}} \frac{|g_{\mathbf{q}}|^2}{\omega_{\mathbf{q}}}$ is the normalized small polaron binding energy. Here, we can separate the lattice distortion into the effective tight-binding model, phonon, and the phonon-polaron interaction Hamiltonian.

With the help of the equation of motion method for the phonon's Green's function (*25*), the theory indicates the phonon frequency dressed by the small polaron will be renormalized:

$$\Omega_{\mathbf{q}} = \omega_{\mathbf{q}}\left(1 + \frac{4|g_{\mathbf{q}}|^2}{\omega_{\mathbf{q}}^2 N}\sum_{\mathbf{k}n}\frac{te^{-S_n}}{\omega_{\mathbf{q}}}(1-\cos\mathbf{q}\cdot\mathbf{R_n})n_{F\mathbf{k}}e^{+i\mathbf{k}\cdot\mathbf{R_n}}\right), \qquad (3)$$

where $S_n = \frac{2}{N}\sum_{\mathbf{q}}\frac{|g_{\mathbf{q}}|^2}{\omega_{\mathbf{q}}^2}\sin^2\left(\frac{\mathbf{q}\cdot\mathbf{R_n}}{2}\right)(2n_{B\mathbf{q}}+1)$. The $e^{-S_n}$ term is the effective factor of electron hopping, which reflects the localization of the small polaron and leads to the reduction of carrier mobility. $n_{F\mathbf{k}}$ and $n_{B\mathbf{q}}$ are the corresponding Fermi-Dirac distribution and Bose-Einstein distribution, which introduce the temperature dependence of the normalized frequency. For polar phonon (which is shown to only exist in thin tellurene), the electron-phonon coupling can develop a long-range interaction, i.e., $g_{\mathbf{q}}^* \propto \frac{1}{q}$, resulting in a finite polaron-phonon coupling when $|g_{\mathbf{q}}|^2$ and $(1-\cos\mathbf{q}\cdot\mathbf{R_n})$ cancel the $\mathbf{q}$ dependence at the $\mathbf{q}\to 0$ limit, and then $\Omega_{\mathbf{q}}$ should have a significant enhancement with strong electron-phonon coupling. For semi-metallic bulk tellurium, dielectric screening only results in weak electron-phonon coupling. Hence, no phonon hardening effect is observed for thick tellurene films corresponding to large polarons (Fig. 3B, blue dots). Whereas few-layer tellurene is semiconducting, the reduced dielectric screening can lead to long-range interactions between polar $A_1$ phonons and electron clouds, contributing to the phonon hardening effect for small polarons (Fig. 3B, red dots). This is consistent with the experimental observation where the phonon frequency is increased drastically by 12 cm$^{-1}$ as the thickness decreases below 10 nm (Fig. 1B). The phonon frequency also exhibits different temperature dependencies when comparing small polarons and large polarons (figs. S6-7). Fig. S7 plots the frequency difference of the $A_1$ phonon between thin and thick tellurene (with thickness below and above the polaron transition). In the experiment, the frequency difference continues to decrease at lower temperatures, which aligns with our theoretical predictions of the frequency difference between small and large polarons.

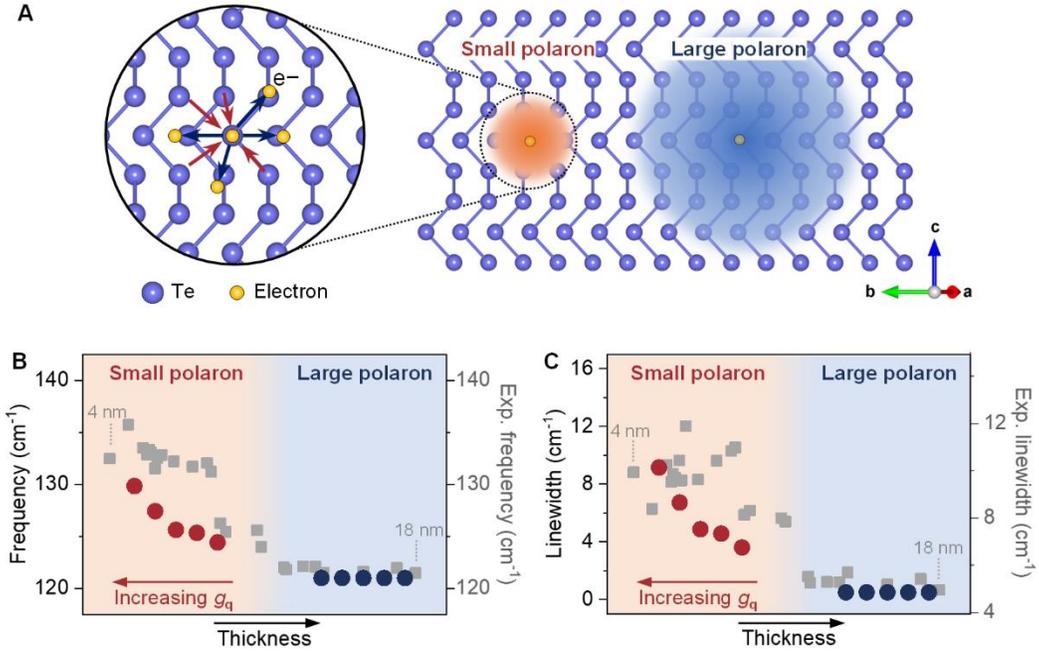

**Fig. 3. Theory-predicted phonon hardening and broadening effect.** (**A**) Illustration of polarons. Left: Arrows represent attractive (red) and repulsive (blue) forces. Right: The red sphere with a small radius and the blue sphere with a large radius represent the formation of the small polaron

and large polaron, respectively. (**B**) The theoretically predicted phonon frequency for small polaron and large polaron. The thickness dependence is schematically sketched. (**C**) The theoretically predicted 1-loop polaron corrected phonon linewidth for small polaron and large polaron. The theoretical linewidth only includes the polaron contribution. The thickness dependence is schematically sketched. The experimental data for tellurene from 4 nm to 18 nm are shown by the closed squares.

To obtain the phonon linewidth, we perform the perturbation of the hopping amplitude and take the 1-loop polaron "polarization operator" correction as an approximation to get the self-energy of a phonon, the phonon linewidth from the small polaron can be derived as (Supplementary Note 2):

$$\Gamma_{\mathbf{q}} = \frac{\omega_{\mathbf{q}}}{\Omega_{\mathbf{q}}} \delta(\omega_{\mathbf{q}} - \Delta) \frac{2\pi e^2 t^2}{\omega_{\mathbf{q}}} \left(\frac{1}{\varepsilon_\infty} - \frac{1}{\varepsilon_0}\right) \sum_n \frac{(\mathbf{q} \cdot \mathbf{R_n})^2}{q^2}, \tag{4}$$

where $\Delta \approx \lim_{\mathbf{q}\to 0} \varepsilon_{\mathbf{p+q}b} - \varepsilon_{\mathbf{p}a}$ is the gap of the polaron band, $a$ and $b$ are band indices as we require multiple polaron bands to describe the semiconducting feature of the few-layer tellurene under the quasiparticle picture. For simplicity, we make use of the two-band model by assuming the initial polaron state is fully occupied, $n_F(\varepsilon_{\mathbf{p}a}) = 1$, and the final state is empty, $n_F(\varepsilon_{\mathbf{p+q}}) = 0$, and the phonon energy is around $\Delta$ to assist the inter-band transition from the initial state to the final state. $\varepsilon_0$ and $\varepsilon_\infty$ are the related dielectric constant. Then the phonon linewidth will exhibit a finite increase due to the influence of small polaron compared to the case of large polaron when $\mathbf{q} \to 0$. As we experimentally observe the increase of linewidth when Te thickness is reduced (Fig. 1E), this increase is also captured from our theory through the 1-loop polaron correction perturbation theory estimation (Fig. 3C). The discrepancy between the absolute linewidth from the experiment and theory can be attributed to the fact that our polaron theory only considers the linewidth broadening due to electron-phonon interaction, while other factors, such as phonon-phonon interaction and defect scattering, can also affect the measured phonon linewidth. Overall, the agreement between the experiment and theory indicates that the enhanced electron-phonon scattering dominates the increase of linewidth for few-layer tellurene. It is also worthwhile mentioning that since small polaron and large polaron theory begin with different canonical bases, a unified theory that treats both, including predicting crossover directly, has not been developed to the best of our knowledge. Nevertheless, our DFT calculation shows strong evidence of the polar $A_1$ phonon, which provides the basis of our picture; therefore, we computed the phonon properties based on small-polaron for thinner films and large-polaron for thicker films, separately.

As a result, our theory can well explain the phonon behaviors observed in few-layer tellurene by phonon renormalization in the small polaron (strong coupling) regime on a semi-quantitative level. All the numerical calculations based on the derived polaron-phonon theory give a minimal estimation with a clear physical picture and align with the experimental observation in Fig. 3. Besides phonon behaviors, the observed sudden drop of carrier mobility (Fig. 1H) due to the formation of small polaron can also be interpreted from a theoretical viewpoint, as the electron-phonon coupling evidenced by the increase of the phonon linewidth will lead to the drop of carrier mobility (Supplementary Note 3). More specifically, the developed small polaron is spatially localized due to the strong electron-phonon scattering. As a result, the effective hopping, which dominates the carrier mobility, is reduced. This further confirms the polaron crossover from large polaron (weak coupling regime) in the bulk tellurium to small polaron (strong coupling regime) in a few-layer tellurene.

Here we comment on the uniqueness of Te in exhibiting such a transition from large polaron to small polaron with thickness reduction. Small polarons require a polar phonon and strong electron-phonon interaction. They have been observed in a number of materials such as transition metal oxides (*26*), organic semiconductors (*27*), 2D materials (*28, 29*), etc. Transitions between large and small polarons have also been observed in perovskite manganese oxides (*12, 13*) like $La_{0.75}Ca_{0.25}MnO_3$ through the metal-semiconductor/insulator transition with the change of doping or temperature. Bulk Te has a non-polar $A_1$ phonon, while this $A_1$ phonon becomes polar with reduced thickness. Such a polar phonon is important in forming strong electron-phonon interaction that leads to small polarons. The transition from non-polar to polar phonon with reduced thickness is due to the change of nearest-neighbor chains of Te. Since Te is a quasi-1D structure as confirmed by magneto-transport and strain-dependent Raman measurements (*19*), the interactions for one chain with adjacent in-plane or out-of-plane chains are equal in bulk tellurium. This is different from conventional 2D materials where the phonon polarity does not change significantly as thickness varies since the in-plane bond is much stronger than the out-of-plane van der Waals bond. Overall, the unique quasi-1D structure of Te and the $A_1$ vibrational profile leads to a change of $A_1$ polarity with different thicknesses, eventually resulting in the transition from large to small polarons and modulation of phonon frequency and linewidth.

**Interchain and intrachain distances show structural changes**

Considering the structural evolution associated with the change of tellurene thickness, an effective method to investigate the atomic arrangements in quasi-1D tellurene with varying thicknesses is through extended X-ray absorption fine structure (EXAFS) analysis. The EXAFS analyzes the oscillatory features in the X-ray absorption spectrum beyond the absorption edge, which reveals the bond distance between the central atom and their neighboring atoms. Fig. 4A shows the normalized x-ray absorption spectra at Te K-edge for tellurene with thicknesses of 18 nm, 14 nm, and 7 nm. The absence of the pre-edge feature is consistent with the X-ray absorption near-edge fine structure of bulk tellurium, indicating that the valence of tellurene remains the same for the analyzed thicknesses (*30*). We then extracted the EXAFS spectra (Fig. 4B) and fitted the bond distance based on the structure model in fig. S8. After Fourier transformation, the EXAFS in Fig. 4C shows a major peak located at around 2.8 Å that is assigned to the Te-Te bond between the nearest atoms. This confirms that there is no phase transition present for tellurene films with different thicknesses. However, the subtle difference in the EXAFS data suggests a slight structural distortion for thinner tellurene films below the critical thickness. Compared to the 14 nm or 18 nm tellurene, tellurene of 7 nm displays an increased amplitude at around 4-5 Å in the R space (Fig. 4C). This suggests an increased scattering of photo-excited electrons by Te atoms in the neighboring chains, indicating the presence of structural changes in tellurene of 7 nm.

The quantitative EXAFS fitting reveals how intrachain and interchain distances vary for different thicknesses (Fig. 4D). For the three thicknesses measured, the average distance between the nearest Te atoms in the same chain, $d_1^{EXAFS}$, is 2.86 Å with a variation of less than 2%. By contrast, the distance between the second-nearest Te atoms in the same chain, $d_2^{EXAFS}$, decreases from 4.76 Å to 4.25 Å for tellurene from 18 nm to 7 nm, demonstrating the contraction of the helical chain due to the lattice deformation. Most notably, the distance between adjacent chains in the lateral direction $d_Y^{EXAFS}$ reduces by 0.19 Å (4%) from 18 nm to 14 nm and further by 0.57 Å (12%) from 14 nm to 7 nm. Contrary to the lateral interchain distance, the distance between chiral chains in the vertical direction shows a slight increase for thinner tellurene of 7 nm as compared to thicker flakes (fig. S8), which is in agreement with our calculations discussed in the next section. The analyzed interchain distances corroborate the structural distortion between neighboring chains promoting the formation of small polarons. As the thickness decreases, tellurium chains in the same plane with respect to the underlying substrate move closer while tellurium chains in the vertical direction displace further away.

The difference in interchain distance between two thicknesses is consistent with the low-frequency phonon modes representing relative vibrations between Te chains. Similar to high-frequency modes, the low-frequency interchain modes also display a strong dependence on tellurene thickness (fig. S9). Below the critical thickness of 10 nm, the interchain modes gain amplitude and blue shift as shown in fig. S9, suggesting an enhanced interchain coupling strength (*31, 32*). This can be potentially explained by the relative displacement between the helical chains in the vertical and horizontal directions observed in the EXAFS fitting. Moreover, the anisotropy of the interchain modes is stronger for thinner tellurene (fig. S10), which is also indicative of an interchain restructuring that leads to a stronger phonon anisotropy.

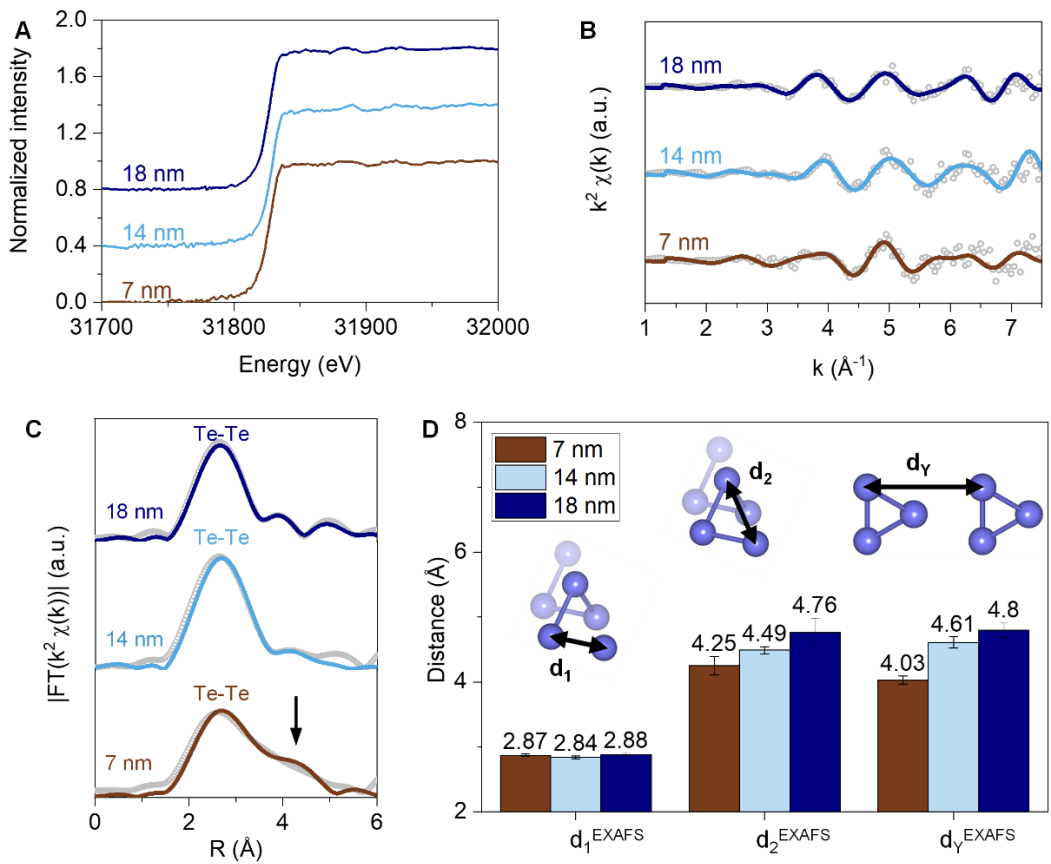

**Fig. 4. X-ray absorption spectra of tellurene for three different thicknesses**. (**A**) Normalized x-ray near-edge fine structure spectra of tellurene for three different thicknesses at Te K-edge. (**B**) The EXAFS (open dots) and its fitting (solid curves) at a $k^2$-weight and with background subtracted. (**C**) Fourier transform of EXAFS data. (**D**) The intrachain distance between the nearest Te atoms $d_1^{EXAFS}$, the distance between the second nearest Te atoms in a single chain $d_2^{EXAFS}$, and the lateral distance between neighboring chains $d_Y^{EXAFS}$ were obtained from the EXAFS fitting.

We then compare the experimental bond distance to the computed crystal structure for few-layer tellurene based on DFT calculations. The calculated atomic distances within the same chain and between adjacent chains were averaged within the unit cell (Figs. 5A-B and fig. S11). The intrachain distance $d_1$ and $d_2$ are reduced by approximately 1% from 6 layers to 2 layers (Figs. 5C-D), consistent with the bond distances obtained from EXAFS fitting. In comparison, the lateral interchain distance shortens for thinner tellurene as shown by the computed lattice constant in Fig. 5E. This is again in agreement with the smaller $d_Y^{EXAFS}$ for the 7 nm tellurene than that of the 18 nm tellurene in the EXAFS measurement. More importantly, the increase of the interchain separation in the vertical direction obtained from EXAFS is reproduced by calculation (Fig. 5D). The lateral contraction and the vertical expansion revealed by the calculations support our EXAFS analysis where few-layer tellurene behaves more like a layered material rather than bulk materials composed of 1D atomic chains. This is in agreement with prior reports where few-layer tellurene can exhibit a covalent-like quasi-bonding (CLQB) between neighboring chains (*23*). Such structural uniqueness contributes to the formation of polar phonons and hence small polarons in few-layer tellurene.

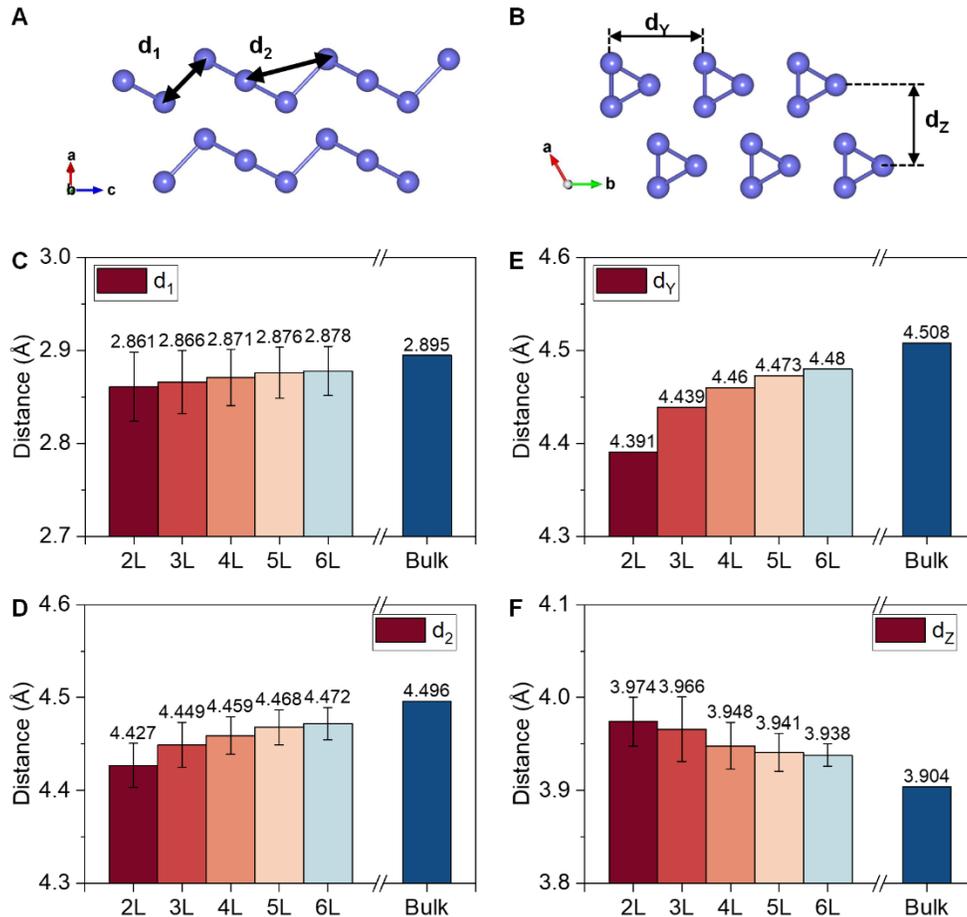

**Fig. 5. Intrachain and interchain distances by first-principles calculations.** (**A**) Schematics of the distance between the nearest Te atoms $d_1$ and the distance between the second nearest Te atoms in a single chain $d_2$. (**B**) Horizontal interchain distance $d_Y$ and vertical interchain distance $d_Z$. (**C-F**) The intrachain distances (**C**) $d_1$ and (**D**) $d_2$, and interchain distances (**E**) $d_Y$ and (**F**) $d_Z$ are computed by first-principles calculations for tellurene with different layer numbers. The error bar represents the standard deviation within one unit cell.

## Discussion

In summary, we investigated the unique behavior of the $A_1$ phonon in quasi-1D tellurene structures of varying thicknesses. A pronounced blue shift and broadening were observed for the $A_1$ phonon as tellurene thins down, indicating a transition from large polarons in bulk tellurium to small polarons in few-layer tellurene. The increasing polar nature of the $A_1$ phonon, predicted by first-principles calculations, further substantiates the formation of small polarons with reduced thickness. This transition from non-polar to polar phonons in quasi-1D tellurene is fundamentally related to changes in the nearest-neighbor chains of tellurene. Unlike conventional 2D materials, where phonon polarity remains stable with varied thicknesses due to strong in-plane bonds, the quasi-1D structure allows for significant modulation in phonon characteristics across thicknesses. This structural evolution influences electron-phonon interactions, leading to the formation of small polarons characterized by localized charge carriers and reduced mobility. Our theoretical model elucidates the phonon hardening and broadening effects due to polaron formation, providing insights into the underlying electron-phonon interactions. We then experimentally and theoretically validated the structural modification in tellurene with a varied number of layers. The EXAFS spectroscopy revealed a reduction in the lateral distance between neighboring helical chains of tellurene atoms for smaller thicknesses, aligning with the crossover between large polarons and small polarons. Our research contributes to a deeper understanding of the relationship between polarons and phonon properties in low-dimensional elemental materials. Future investigations could explore the implications of small polaron formation on device performance and the potential of tellurene-based materials for advanced functional devices.

## Materials and Methods

**Raman spectroscopy:** The Raman spectra were measured on a Horiba LabRAM spectrometer at room temperature, which was equipped with an ultra-low-frequency filter module for low-frequency measurement. The 532 nm laser beam was incident and collected through a 100× objective with a spot size of 1 μm². To avoid beam damage, the laser power was kept below 40 μW. The polarization-resolved measurement was performed with a polarizer added to the detection path of the spectrometer. The differential reflectance spectra were measured with the Laser-Driven Light Sources as the broadband white light.

**X-ray absorption spectroscopy:** Extended X-ray Absorption Fine Structure (EXAFS) analysis was performed on tellurene to investigate their local atomic structure. The tellurene flakes of varied thickness were deposited on Kapton film. EXAFS measurements were conducted at the Te K-edge (31.814 keV) using synchrotron radiation at 20-BM at the Advanced Photon Source (APS), Argonne National Laboratory. The incident x-ray beam of 500 × 500 μm² was collimated to approximately 20 × 20 μm² and the samples were scanned to find individual tellurene flakes. The monochromator was detuned by 15% to reduce x-rays with higher harmonic energies. EXAFS spectra were collected in transmission mode using ionization chambers filled with argon gas. The spectra were analyzed using the Athena and Artemis software packages (*33*). The absorption spectra were processed by removing the pre-edge and normalized to the edge step, followed by standard background removal and Fourier transformed to R-space for EXAFS modeling. The ultrathin Te samples make it very challenging to obtain high data quality at high $k$ space. The detailed fitting results can be found in Supplementary Table 1.

**Analytical Theory:** The renormalized phonon frequency could be calculated through the equation of motion method for the phonon's Green's function. If we perform the perturbation of the hopping amplitude under the proper assumption, we can simplify the Hamiltonian and calculate the phonon linewidth based on the 1-loop polaron correction. The detailed derivation of the comprehensive analytical theory can be found in the Supplementary Materials.

**First-principles calculations:** First-principles calculations were carried out using the Vienna *ab initio* simulation package (VASP v.5.4.4) with projector augmented wave (PAW) pseudopotentials for electron-ion interactions (*34*) and the Perdew-Burke-Ernzerhof (PBE) functional for exchange-correlation interactions (*35*). For bulk Te, both atoms and cell volume were allowed to relax until the residual forces were below 0.001 eV/Å, with a cutoff energy set at 300 eV and a gamma-centered 18×18×14 k-point sampling. The optimized lattice constants were $a = b = 4.508$ Å, and $c = 5.959$ Å, where the chiral Te chain direction is along the $c$ direction (see Fig. 1A). Few-layer Te systems were then modeled by a periodic slab geometry, where a vacuum separation of 22 Å in the out-of-plane direction was used to avoid spurious interactions with periodic images. As a common practice, the out-of-plane direction is defined as the $z$ direction. For the 2D slab calculations, the chiral Te chain direction is defined as the $x$ direction while the inter-chain direction is along the $y$ axis, where 14×18×1 k-point samplings were used. All atoms were relaxed until the residual forces were below 0.001 eV/Å and the in-plane lattice constants were also optimized. Phonon calculations were performed using the optimized structures. The dynamic matrix was calculated using the finite difference scheme implemented in the Phonopy software (*36*). Hellmann-Feynman forces in the supercell (3×3×2 for the bulk and 2×3×1 for few-layer systems) were computed by VASP for both positive and negative atomic displacements ($\delta = 0.03$ Å) and then used in Phonopy to construct the dynamic matrix, whose diagonalization provides phonon frequencies and phonon eigenvectors (i.e., vibrations). After obtaining the eigenvectors of phonon modes in the bulk and few-layer Te systems, we then calculated the change of the dipole moment introduced by a phonon vibration. For the system in equilibrium, we computed the electric dipole moment using the Berry phase method (*37*), and then introduced perturbation into the atomic structure by the amount of the phonon eigendisplacements (i.e., phonon eigenvectors normalized by the atomic mass) and re-computed the dipole moment. The difference of the dipole moment between the equilibrium and perturbed structures is thus the change of the dipole moment by the phonon vibration.

**Acknowledgments**

**Funding:** K.Z. and S.H. acknowledge the support from the National Science Foundation (NSF) (Grant Nos. ECCS-2246564, ECCS-1943895, ECCS-2230400, and DMR-2329111), Air Force Office of Scientific Research (AFOSR) under grant FA9550-22-1-0408, and Welch Foundation (Award No. C-2144). C.F. acknowledges support from the U.S. Department of Energy (DOE), Office of Science (SC), Basic Energy Sciences (BES), Award No. DE-SC0020148, while M.L. thanks support from the National Science Foundation (NSF) Designing Materials to Revolutionize and Engineer our Future (DMREF) Program with Award No. DMR-2118448. W. W. Z. acknowledges support from Air Force Office of Scientific Research under award number FA2386-21-1-4064. The synthesis of tellurene was supported by NSF under grant No. CMMI-2046936. This research used resources of the Advanced Photon Source, a U.S. Department of Energy (DOE) Office of Science user facility at Argonne National Laboratory and is based on research supported by the U.S. DOE Office of Science-Basic Energy Sciences, under Contract No. DE-AC02-06CH11357. A portion of this research (DFT calculations) used resources at the Center for Nanophase Materials Sciences, which is a U.S. Department of Energy Office of Science User Facility. This work was partly supported by the U.S. Department of Energy, Office of Science, Office of Basic Energy Sciences, Materials Sciences and Engineering Division (M.Y.) and by the U.S. Department of Energy (DOE), Office of Science, National Quantum Information Science Research Centers, Quantum Science Center (S.-H.K.). This research also used resources of the National Energy Research Scientific Computing Center, a DOE Office of Science User Facility supported by the Office of Science of the U.S. Department of Energy under Contract No. DE-AC02-05CH11231 and using NERSC award BES-ERCAP0024568. L.L. acknowledges computational resources of the Compute and Data Environment for Science (CADES) at the Oak

Ridge National Laboratory, which is supported by the Office of Science of the U.S. Department of Energy under Contract No. DE-AC05-00OR22725.

**Author contributions:**
Conceptualization: KZ, SH
Optical spectroscopy and analysis: KZ
Theory: CF and PS
EXAFS experiments and analysis: KZ, SK, GW
DFT calculations: LL, SK
Sample synthesis and preparation: JJ, RZ, YW
Supervision: MY, PY, WW, ML, SH
Writing–original draft: KZ, CF
Writing–review and editing: KZ, CF, ML, SH

**Competing interests:** Authors declare that they have no competing interests.

**Data and materials availability:** All data are available in the main text or the supplementary materials.

**Supplementary Text**

**Supplementary Note 1.** Breit-Wigner-Fano (BWF) line shape.

The Breit-Wigner-Fano (BWF) line shape is defined in this paper as

$$I(\omega) = I_0 \frac{[1 + (\omega - \omega_0)/(q_s \Gamma_0/2)]^2}{1 + [(\omega - \omega_0)/(\Gamma_0/2)]^2} \qquad (1)$$

in which $I_0$, $\omega_0$, and $\Gamma_0$ denote the intensity, frequency, and linewidth of the modified discrete state, respectively. The asymmetry parameter $1/q_s$ quantifies the level of Fano resonance.

**Supplementary Note 2.** Polaron theory.

We begin with the general electron-phonon interacting Hamiltonian, assuming spinless electrons and also single-mode phonons with normalized $\hbar = 1$,

$$H = H_{0e} + H_{0ph} + H_{e-ph} = -t\sum_{\langle ij \rangle} c_i^\dagger c_j + \sum_{\mathbf{q}} \omega_{\mathbf{q}} a_{\mathbf{q}}^\dagger a_{\mathbf{q}} + \sum_{j\mathbf{q}} g_{\mathbf{q}} c_j^\dagger c_j (a_{\mathbf{q}} + a_{-\mathbf{q}}^\dagger) e^{i\mathbf{q}\cdot\mathbf{R}_j}, \quad (5)$$

$g_{\mathbf{q}}$ is the electron-phonon coupling constant where the detailed form depends on the type of electron-phonon interactions (*1*), the notation $\langle ij \rangle$ indicates that sites $i$ and $j$ are nearest neighbors. $H_0 = H_{0e} + H_{0ph}$ is the non-interacting Hamiltonian for electron and phonon, respectively. $H_{e-ph}$ is the electron-phonon interaction term. This model describes the coupling between local polarization at site $j$ with the electron density.

**Large polaron model:**

For weak coupling between electrons and phonons, with large electron bandwidth, we can solve it in momentum space. Defining $c_{\mathbf{k}} = \frac{1}{\sqrt{N}} \sum_j c_j e^{-i\mathbf{k}\cdot\mathbf{R}_j}$, we have

$$H_0 = -t \sum_{\mathbf{k}} W_{\mathbf{k}} c_{\mathbf{k}}^+ c_{\mathbf{k}} + \sum_{\mathbf{q}} \omega_{\mathbf{q}} \left( a_{\mathbf{q}}^+ a_{\mathbf{q}} + \frac{1}{2} \right), \quad (6)$$

$$H_{\text{e-ph}} = \sum_{\mathbf{kq}} g_{\mathbf{q}} c_{\mathbf{k+q}}^+ c_{\mathbf{k}} (a_{\mathbf{q}} + a_{-\mathbf{q}}^+), \quad (7)$$

where $W_{\mathbf{k}} = \sum_{NN} e^{i\mathbf{k}\cdot\mathbf{R}_{NN}}$, and defining $\varepsilon_{\mathbf{k}} = tW_{\mathbf{k}}$ and $\mathbf{R}_{NN} = \mathbf{R}_i - \mathbf{R}_j$ is the relative distance for nearest neighbors. Defining the phonon propagator as

$$D(\mathbf{q},\tau) = -T_\tau \left\langle \left(a_{\mathbf{q}}^+(\tau) + a_{-\mathbf{q}}(\tau)\right)\left(a_{-\mathbf{q}}^+(0) + a_{\mathbf{q}}(0)\right)\right\rangle; D(\mathbf{q}, i\omega_n) = \int_0^\beta D(\mathbf{q},\tau) e^{+i\omega_n \tau} d\tau. \quad (8)$$

For non-interacting phonons, we have $D^{(0)}(\mathbf{q}, i\omega_n) = \frac{2\omega_{\mathbf{q}}}{(i\omega_n)^2 - \omega_{\mathbf{q}}^2}$. With electron-phonon interaction, the polarization operator can be written as

$$\Pi^{(1)}(\mathbf{q}, i\omega_n) = \frac{1}{\beta V} \sum_{\mathbf{p}, ip_n} G^{(0)}(\mathbf{p}, ip_n) G^{(0)}(\mathbf{p}+\mathbf{q}, ip_n + i\omega_n). \quad (9)$$

The renormalized phonon Green's function can be written as

$$D(\mathbf{q}, i\omega_n) = \frac{2\omega_{\mathbf{q}}}{(i\omega_n)^2 - \omega_{\mathbf{q}}^2 - 2\omega_{\mathbf{q}}(g_{\mathbf{q}})^2 \Pi^{(1)}(\mathbf{q}, i\omega_n)}. \quad (10)$$

The renormalized phonon frequency and lifetime can be solved as the poles,

$$\left(\Omega_{\mathbf{q}} - i\Gamma_{\mathbf{q}}\right)^2 = \omega_{\mathbf{q}}^2 + 2\omega_{\mathbf{q}} |g_{\mathbf{q}}|^2 \Pi^{(1)}(\mathbf{q}, i\omega_n), \quad (11)$$

which, given the small electron-phonon coupling constant $g_{\mathbf{q}}$, the modulation to phonon frequency in the weak coupling regime is negligible. While the phonon linewidth under condition $\mathbf{q} \to 0$, we have $\Gamma_0 = -\frac{\omega_0}{\Omega_0} |g_{\mathbf{q}}|^2 \text{Im}\Pi^{(1)}(\mathbf{q} \to 0, \omega_0) \to 0$, $\omega_0$ is the unnormalized frequency when $\mathbf{q} \to 0$.

**Small polaron model:**

After performing the small polaron where the polaron is more localized in real space and electrons are strongly coupled to the phonons, the so-called Lang-Firsov unitary transformation (*1*), finally the Hamiltonian after canonical transform can be rewritten as

$$H = -\sum_j E_B c_j^+ c_j - t \sum_{\langle ij \rangle} exp\left(-\sum_\mathbf{q} T_{\mathbf{q}ij} p_\mathbf{q} e^{-i\mathbf{q}\cdot\mathbf{R}_i}\right) c_i^+ c_j + \sum_\mathbf{q} \omega_\mathbf{q} a_\mathbf{q}^+ a_\mathbf{q}, \qquad (12)$$

where the canonical phonon operators are defined in convention as $u_\mathbf{q} = \sqrt{\frac{\hbar}{2m\omega_\mathbf{q}}}(a_\mathbf{q} + a_{-\mathbf{q}}^+)$ and $p_\mathbf{q} = i\sqrt{\frac{\hbar m \omega_\mathbf{q}}{2}}(a_\mathbf{q}^+ - a_{-\mathbf{q}})$, $T_{\mathbf{q}ij} = \frac{g_\mathbf{q}^*}{\omega_\mathbf{q}} i \sqrt{\frac{2}{\hbar m \omega_\mathbf{q}}}(1 - e^{i\mathbf{q}\cdot(\mathbf{R}_i - \mathbf{R}_j)})$, $E_B = \sum_\mathbf{q} \frac{|g_\mathbf{q}|^2}{\omega_\mathbf{q}}$ is the normalized small polaron binding energy. The small polaron can lead to phonon hardening, which is shown in Ref. (*2*) for instance. Through the standard procedure of the equation of motion for the phonon Green's function, the renormalized phonon frequency can be written as

$$\Omega_\mathbf{q} = \omega_\mathbf{q}\left(1 + \frac{4|g_\mathbf{q}|^2}{\omega_\mathbf{q}^2 N}\sum_{\mathbf{k}n}\frac{t e^{-S_n}}{\omega_\mathbf{q}}(1 - \cos\mathbf{q}\cdot\mathbf{R}_\mathbf{n}) n_{F\mathbf{k}} e^{+i\mathbf{k}\cdot\mathbf{R}_\mathbf{n}}\right), \qquad (13)$$

where $S_n = \frac{2}{N}\sum_\mathbf{q}\frac{|g_\mathbf{q}|^2}{\omega_\mathbf{q}^2}\sin^2\left(\frac{\mathbf{q}\cdot\mathbf{R}_\mathbf{n}}{2}\right)(2n_{B\mathbf{q}} + 1)$. $e^{-S_n}$ is the effective factor of the electron hopping, which reflects the effective localization of the small polaron and leads to the reduction of the electrical mobility. Based on Holstein's original estimation (*3*), the range of the value for $e^{-S_n}$ is $10^{-2} \sim 10^{-4}$, we take $e^{-S_n} = 10^{-3}$ during the numerical calculation. Notice that the effective hopping factor $e^{-S_n}$ also have the temperature dependence. However, the non-interacting phonon frequency before the normalization also has an approximately linear temperature dependence indicated by the experimental data, this makes the factor in the Boson statistics, $\frac{\hbar\omega}{kT}$, becomes weakly dependent on the temperature. As a result, we assume the value of the effective hopping factor remains relatively stable around room temperature. Besides, applying the linear temperature dependence for factor $S$ will cause a large change in the frequency renormalization due to the exponential behavior, which violates the experimental observation of the temperature dependence of the phonon frequency.

For the polar phonon, we take the form of $g_\mathbf{q} = \frac{\sqrt{2\pi e^2 \omega_\mathbf{q}\left(\frac{1}{\epsilon_\infty} - \frac{1}{\epsilon_0}\right)}}{\mathbf{q}}$ (*1*), and $e$ is the elementary charge. From the previous literature (*4*), we can get the layer-dependent dielectric constant, and we take the empirical approximation that $\varepsilon_\infty = 0.5\varepsilon_0$, here $\varepsilon_0$ and $\varepsilon_\infty$ are dielectric constant for static case and infinity frequency, respectively. The static dielectric constant of two to six layers along the $E \perp C$ direction from the literature mentioned above is taken as 10.3, 14.3, 20, 21.4, and 27.3, respectively [4]. The phonon frequency simulated using the static dielectric constant along the $E \parallel C$ direction is consistent with the results based on the static dielectric constant along the $E \perp C$ direction. We set the chemical potential as 0.75 eV based on the literature (*5, 6*), assuming the electron hopping as $t = 0.21\hbar\omega_\mathbf{q}(\mathbf{q} \to 0)$ and assuming $\omega_\mathbf{q}(\mathbf{q} \to 0) = 121$ cm$^{-1}$ for the A$_1$ phonon, with the first nearest neighbor distance being 0.289 nm based on our EXAFS measurement presented in **Figure S8**, we finally can get the numerically calculated results presented in the main content. For the temperature-dependent frequency calculations in **Figure S7** (b), the static dielectric constant is taken as 14.3, and the phonon frequency before the normalization $\omega_\mathbf{q}$ is assumed to be linearly dependent on the temperature from 125.84 cm$^{-1}$ to 121 cm$^{-1}$ within the temperature range from 80K to 300K. If we set $F_n = \frac{4|g_\mathbf{q}|^2}{\omega_\mathbf{q}^2 N}\sum_\mathbf{k}\frac{t e^{-S_n}}{\omega_\mathbf{q}}(1 - \cos\mathbf{q}\cdot\mathbf{R}_\mathbf{n})n_{F\mathbf{k}}e^{+i\mathbf{k}\cdot\mathbf{R}_\mathbf{n}}$ as the phonon hardening factor, we can also observe a dependence of phonon frequency with various factors such as the static dielectric constant, the nearest neighbor distance, and the chemical potential (see **Figure S12**). The static dielectric constant for (b) and (c) is assumed as 8, and other parameters are taken as mentioned before.

Then we perform the perturbation of the hopping amplitude, $t_{ij} = t \exp\left(-\sum_\mathbf{q} T_{\mathbf{q}ij} p_\mathbf{q} e^{-i\mathbf{q}\cdot\mathbf{R}_i}\right) \approx t\left(1 - \sum_\mathbf{q} T_{\mathbf{q}ij} p_\mathbf{q} e^{-i\mathbf{q}\cdot\mathbf{R}_i}\right)$, which is valid when $-\sum_\mathbf{q} T_{\mathbf{q}ij} p_\mathbf{q} e^{-i\mathbf{q}\cdot\mathbf{R}_i}$ is sufficiently small. This should be satisfied as in large $\mathbf{q}$, $g_\mathbf{q}^*$ should have the $1/q$ dependence and will lead to sufficient small $g_\mathbf{q}^*$. For small $\mathbf{q}$, the $1/q$ dependence will cancel with the $\mathbf{q}$ dependence from $(1 - e^{i\mathbf{q}\cdot\mathbf{R}_i})$ within the $T_{\mathbf{q}ij}$. Besides, as we are taking the nearest neighbor, we can assume $\mathbf{R}_i - \mathbf{R}_j$ is small enough to lead to valid perturbation expansion.

Instead of the lattice displacement $u_\mathbf{q}$ and strength $g_\mathbf{q}$ in conventional electron-phonon coupling, the electrons in this regime couple with lattice momentum $p_\mathbf{q}$ with strength $\frac{g_\mathbf{q}^*}{\omega_\mathbf{q}}$. It is feasible to calculate the polaron "polarization" operator that renormalizes phonon energy and gains phonon the lifetime, through the phonon momentum-momentum correlator. Performing the Fourier transform $c_i = \frac{1}{\sqrt{N}} \sum_\mathbf{k} e^{-i\mathbf{k}\cdot\mathbf{R}_i} c_\mathbf{k}$, the Hamiltonian becomes

$$H' = -\sum_\mathbf{k} \left(E_B + t \sum_n e^{i\mathbf{k}\cdot\mathbf{R}_n}\right) c_\mathbf{k}^+ c_\mathbf{k} - t \sum_{\mathbf{kq}} g_{\mathbf{q},\mathbf{k},eff} c_{\mathbf{k}+\mathbf{q}}^+ c_\mathbf{k} \left(a_\mathbf{q}^+ - a_{-\mathbf{q}}\right) + \sum_\mathbf{q} \omega_\mathbf{q} a_\mathbf{q}^+ a_\mathbf{q}, \quad (14)$$

where "n" means the summation of the nearest neighbors, $\mathbf{R}_n$ is the relative distance between two nearest neighbors, and $g_{\mathbf{q},\mathbf{k},eff} = \sum_n \frac{g_\mathbf{q}^*}{\omega_\mathbf{q}} \left(1 - e^{i\mathbf{q}\cdot\mathbf{R}_n}\right) e^{i\mathbf{k}\cdot\mathbf{R}_n}$.

To proceed, we define the dimensionless momentum self-correlation as

$$F(\mathbf{q}, \tau) = -T_\tau \left\langle \left(a_\mathbf{q}^+(\tau) - a_{-\mathbf{q}}(\tau)\right)\left(a_{-\mathbf{q}}^+(0) - a_\mathbf{q}(0)\right)\right\rangle; F(\mathbf{q}, i\omega_n) = \int_0^\beta F(\mathbf{q}, \tau) e^{+i\omega_n \tau} d\tau. \quad (15)$$

Then, using the fact that $a_\mathbf{q}(\tau) = a_\mathbf{q} e^{-\tau \omega_\mathbf{q}}$, we have $F^{(0)}(\mathbf{q}, i\omega_n) = -\frac{2\omega_\mathbf{q}}{(i\omega_n)^2 - \omega_\mathbf{q}^2} = -D^{(0)}(\mathbf{q}, i\omega_n)$. The free-polaron propagator can be written as $G^{(0)}(\mathbf{p}, ip_m) = \frac{1}{ip_m - (E_B + t \sum_n e^{i\mathbf{p}\cdot\mathbf{R}_n})}$, and the 1-loop polaron "polarization operator" can be written as

$$\Pi^{(1)}(\mathbf{q}, i\omega_n) = \frac{t^2}{\beta V} \sum_{\mathbf{p}, ip_n} G^{(0)}(\mathbf{p}, ip_n) G^{(0)}(\mathbf{p} + \mathbf{q}, ip_n + i\omega_n)$$

$$= \frac{t^2}{V} \sum_\mathbf{p} \frac{n_F(\varepsilon_\mathbf{p}) - n_F(\varepsilon_{\mathbf{p}+\mathbf{q}})}{i\omega_n + \varepsilon_\mathbf{p} - \varepsilon_{\mathbf{p}+\mathbf{q}}} g_{\mathbf{q},\mathbf{p},eff} g_{\mathbf{q},\mathbf{p}+\mathbf{q},eff}^*, \quad (16)$$

where $\varepsilon_\mathbf{p} = E_B + t\sum_n e^{i\mathbf{p}\cdot\mathbf{R}_n}$ is the effective polaron eigenenergy. The renormalized phonon Green's function can be written based on the momentum-momentum Green's function as

$$F(\mathbf{q}, i\omega_n) = -\frac{2\omega_\mathbf{q}}{(i\omega_n)^2 - \omega_\mathbf{q}^2 - 2\omega_\mathbf{q} \Pi^{(1)}(\mathbf{q}, i\omega_n)}. \quad (17)$$

The renormalized phonon frequency and lifetime can be solved as the poles. For thicker films, the band structure calculations indicate more semi-metallic behaviors, which has a larger dielectric screening effect. In this regime, we have $g_\mathbf{q}^* \propto$ constant, therefore, $g_{\mathbf{q}\to 0, \mathbf{k}, eff} = \sum_n \frac{g_\mathbf{q}^*}{\omega_\mathbf{q}}\left(1 - e^{i\mathbf{q}\cdot\mathbf{R}_n}\right) e^{i\mathbf{k}\cdot\mathbf{R}_n} \to 0$, since the $1 - e^{i\mathbf{q}\cdot\mathbf{R}_n}$ term will be zero at the wavevector $\mathbf{q} \to 0$. Therefore, a key feature of the polaron-induced phonon hardening is the lack of the screening effect. The thinner film is semiconducting, contributing to less dielectric screening.

For Raman studies, we can just focus on one specific branch of phonon frequency $\omega_\mathbf{q} = \omega_0$, then it appears that $g_{\mathbf{q}\to 0,\mathbf{k},eff} = \sum_n \frac{g_\mathbf{q}^*}{\omega_\mathbf{q}}(1 - e^{i\mathbf{q}\cdot\mathbf{R_n}})e^{i\mathbf{k}\cdot\mathbf{R_n}} \to 0$. However, there is one exception. If there is a polar phonon developed, then the electron-phonon coupling can develop a long-range interaction, i.e., $g_\mathbf{q}^* \propto \frac{1}{q}$, resulting in a finite polaron-phonon coupling at the $\mathbf{q} \to 0$ limit. In this case, the effective phonon-polaron coupling constant can be written as (*1*)

$$g_{\mathbf{q}\to 0,\mathbf{k},eff} = \lim_{\mathbf{q}\to 0}\sum_n \sqrt{\frac{2\pi e^2}{\omega_0}\left(\frac{1}{\varepsilon_\infty} - \frac{1}{\varepsilon_0}\right)}\frac{i(\mathbf{q}\cdot\mathbf{R_n})}{q}e^{i\mathbf{k}\cdot\mathbf{R_n}}. \tag{18}$$

Then, for $\mathbf{q} \to 0$ optical phonon, since we are no longer studying metals but more semiconductors, we can generalize from single-band to multi-band, i.e.,

$$\Pi^{(1)}(\mathbf{q}, i\omega_n) = \frac{t^2}{V}\sum_{\mathbf{p}ab} \frac{n_F(\varepsilon_{\mathbf{p}a}) - n_F(\varepsilon_{\mathbf{p+q}b})}{i\omega_n + \varepsilon_{\mathbf{p}a} - \varepsilon_{\mathbf{p+q}b}} g_{\mathbf{q},\mathbf{p},a,eff}g_{\mathbf{q},\mathbf{p+q},b,eff}^*, \tag{19}$$

where *a* and *b* are band indices. For simplicity, we just use a two-band model. Assuming the initial polaron state is fully occupied $n_F(\varepsilon_{\mathbf{p}a}) = 1$, and the final state is empty $n_F(\varepsilon_{\mathbf{p+q}}) = 0$, and the phonon energy is $\lim_{\mathbf{q}\to 0}\varepsilon_{\mathbf{p+q}b} - \varepsilon_{\mathbf{p}a} \approx \Delta$ as the gap of the polaron band (from occupied to empty band with the assistance of phonons), then we have

$$\Pi^{(1)}(\mathbf{q}\to 0, \omega_0) = \frac{t^2}{V}\sum_\mathbf{p} \frac{g_{0,\mathbf{p},eff}g_{0,\mathbf{p},eff}^*}{\omega_0 - \Delta + i0^+}, \tag{20}$$

from which we obtain that

$$\mathrm{Im}\Pi^{(1)}(\mathbf{q}\to 0, \omega_0) = -\pi\delta(\omega_0 - \Delta)\frac{2\pi e^2 t^2}{\omega_0}\left(\frac{1}{\varepsilon_\infty} - \frac{1}{\varepsilon_0}\right)\sum_n \frac{(\mathbf{q}\cdot\mathbf{R_n})^2}{q^2}. \tag{21}$$

Finally, the phonon linewidth can be written as

$$\Gamma_0 = -\frac{\omega_0}{\Omega_0}\mathrm{Im}\Pi^{(1)}(\mathbf{q}\to 0, \omega_0) = \frac{\omega_0}{\Omega_0}\delta(\omega_0 - \Delta)\frac{2\pi e^2 t^2}{\omega_0}\left(\frac{1}{\varepsilon_\infty} - \frac{1}{\varepsilon_0}\right)\sum_n \frac{(\mathbf{q}\cdot\mathbf{R_n})^2}{q^2}. \tag{22}$$

The 1-loop polaron correction diagram contribution will dominate the contribution of the increase of the phonon linewidth, and this gives the results presented in the main text with the parameters mentioned above, and the large polaron linewidth is set to the value of 0.5 cm$^{-1}$. It is worth mentioning that this polaron gap is not the electronic bandgap. The physical picture under this result indicates that since elemental Te has 3 atoms per unit cell, 2 atoms form the conduction band (CB) and valance band (VB), and the third atom is closer to VB and has a small energy split, assisting the inter-band transition (energy of the A$_1$ phonon around 15meV) between the two bands.

**Supplementary Note 3.** The theoretical explanation for changes in mobility.

We present the theoretical explanation of the drop in mobility in thinner samples qualitatively. Since the electron mobility $\mu$ can be expressed through the electrical conductivity $\sigma$:

$$\sigma = \frac{n_0 e^2 \tau}{m_e} = e\mu n_0, \tag{23}$$

where $n_0$ is the electron density, $e$ is the elementary charge, $m_e$ is the electron mass, $\tau$ is the relaxation time. As a result, we should notice that mobility is proportional to the relaxation time. Remind of the Matthiessen's rule, $\frac{1}{\tau} = \frac{1}{\tau_{ele}} + \frac{1}{\tau_{ph}} + \frac{1}{\tau_{defect}} + \cdots$, and $\tau_{ele}$ is the relaxation time caused by electron-electron scattering, $\tau_{defect}$ comes from the defect scattering, while $\tau_{ph}$ is related to the electron-phonon scattering and should be related to the phonon linewidth: $\frac{1}{\tau_{ph}} \propto \Gamma_0$.

We can see when the thickness of the Te sample is reduced, the increase of the phonon linewidth is confirmed in theory, which leads to the drop in mobility. However, the drop in mobility cannot be attributed solely to the contribution of phonons, even though this theoretical qualitative analysis from the increase of the phonon linewidth matches experimental measurements. How other factors affect mobility is beyond the scope of this theory.

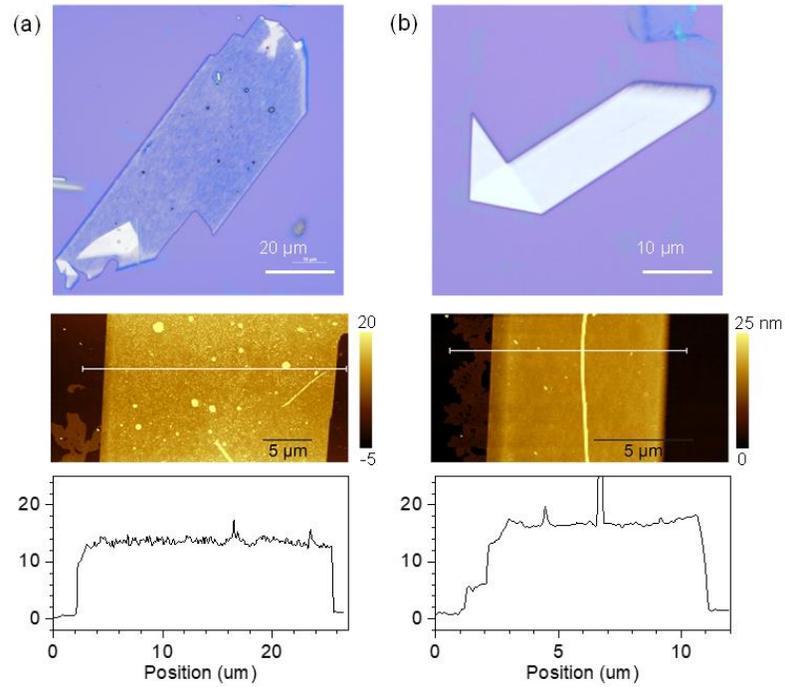

**Fig. S1**. Tellurene with different thicknesses. (a) Thickness of 13.0 nm. (b) Thickness of 15.8 nm. Top panel: optical images. Middle panel: atomic force microscopy (AFM) images. Bottom panel: Line scans across the tellurene thin film.

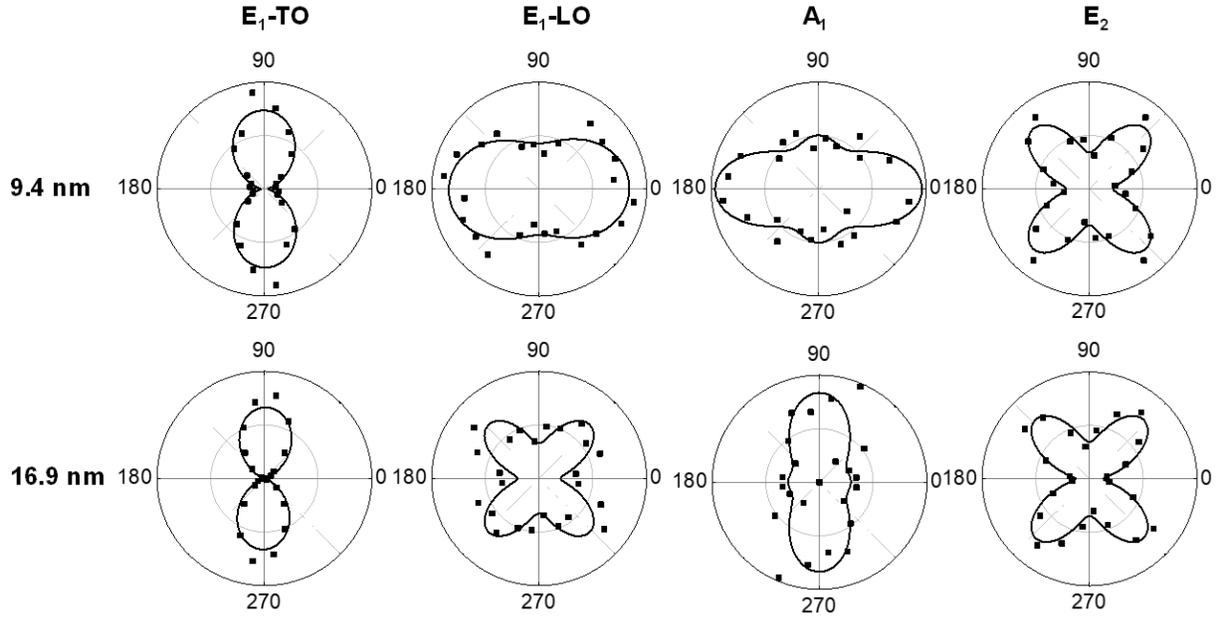

**Fig. S2**. Polar plots of Raman intensity for tellurene of 9.4 and 16.9 nm including the $E_1$-TO mode, $E_1$-LO mode, $A_1$ mode, and $E_2$ mode. For the polarization-dependent measurement, the sample was rotated in-plane, and the scattering that is polarized parallel to the incident light was collected. The polarization angle is defined as the angle between the laser polarization and the [0001] direction. Depending on the phonon symmetry of the $D_3$ point group, the intensity of the characteristic Raman modes changes as a function of the polarization angle. For tellurene with a thickness of 9.1 nm, the $A_1$ mode exhibits a two-fold symmetry with a maximum intensity at 0° and 180° along the [1000] direction. However, the maximum intensity of the $A_1$ mode shifts to 90° and 270° when the thickness of tellurene increases to 16.9 nm, indicating changes in the phonon anisotropy. The different anisotropic dependencies can be well explained by the Raman tensor theory. Based on the Raman tensor, the intensity of the $A_1$ mode can be written as $|\boldsymbol{a}\sin^2\theta + \boldsymbol{b}\cos^2\theta|^2$ in which $\theta$ is the polarization angle and $\boldsymbol{a}$ and $\boldsymbol{b}$ are Raman tensor elements. Using this equation, the polar plots of the 9.4 nm tellurene can be fitted with an $\boldsymbol{a}/\boldsymbol{b}$ ratio of 0.86, while the polar plots of the 16.9 nm tellurene can be reproduced with an $\boldsymbol{a}/\boldsymbol{b}$ ratio of 1.18. The change in the $\boldsymbol{a}/\boldsymbol{b}$ ratio directly accounts for the observation that the maximum Raman intensity of the $A_1$ mode is located at 0° and 180° or 90° and 270° for different thicknesses.

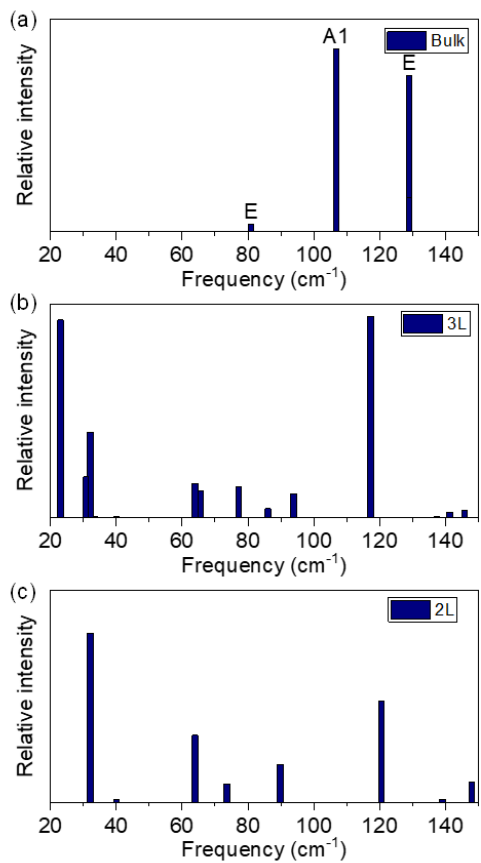

**Fig. S3**. The calculated Raman intensity for (a) bulk tellurium, (b) tellurene composed of three layers of helical chains, and (c) tellurene composed of two layers of helical chains. The tellurene composed of three layers (3L) is predicted to have interchain vibrations at 23.1 cm$^{-1}$ and 30.8 cm$^{-1}$ with relatively strong vibrational amplitudes. While the interchain vibration at 32.1 cm$^{-1}$ is evident for tellurene with two layers (2L).

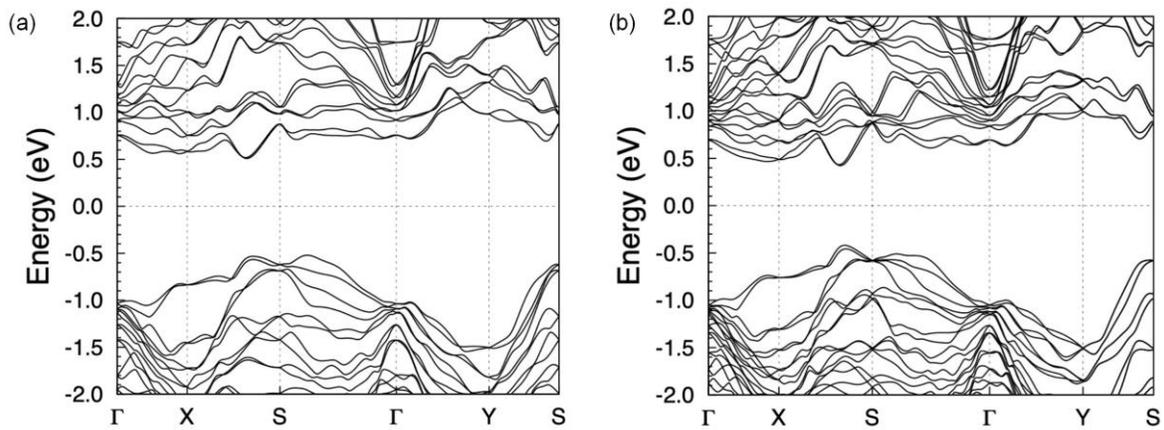

**Fig. S4**. The calculated band structure of (a) 3L and (b) 4L tellurene.

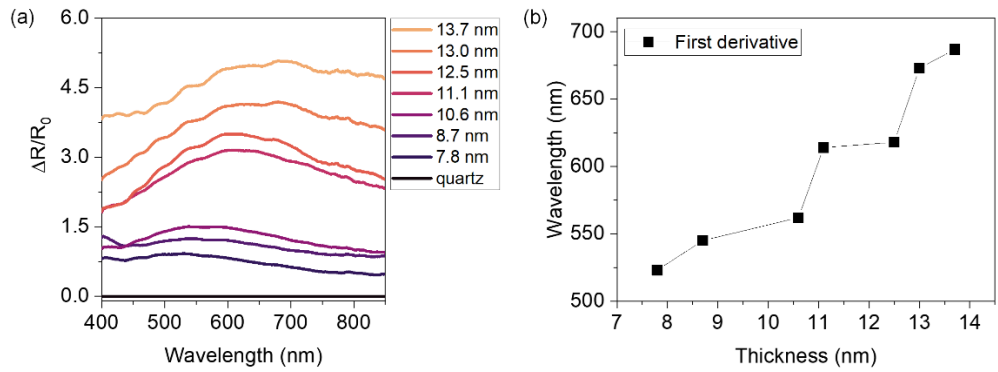

**Fig. S5**. Differential reflectance of few-layer tellurene on quartz. (a) Differential reflectance spectra of tellurene. $\Delta R/R_0$ is defined as the normalized reflectance difference of tellurene/quartz with respect to quartz. (b) The wavelength where the first derivative of differential reflectance equals zero. This indicates the absorption wavelength of tellurene of different thicknesses.

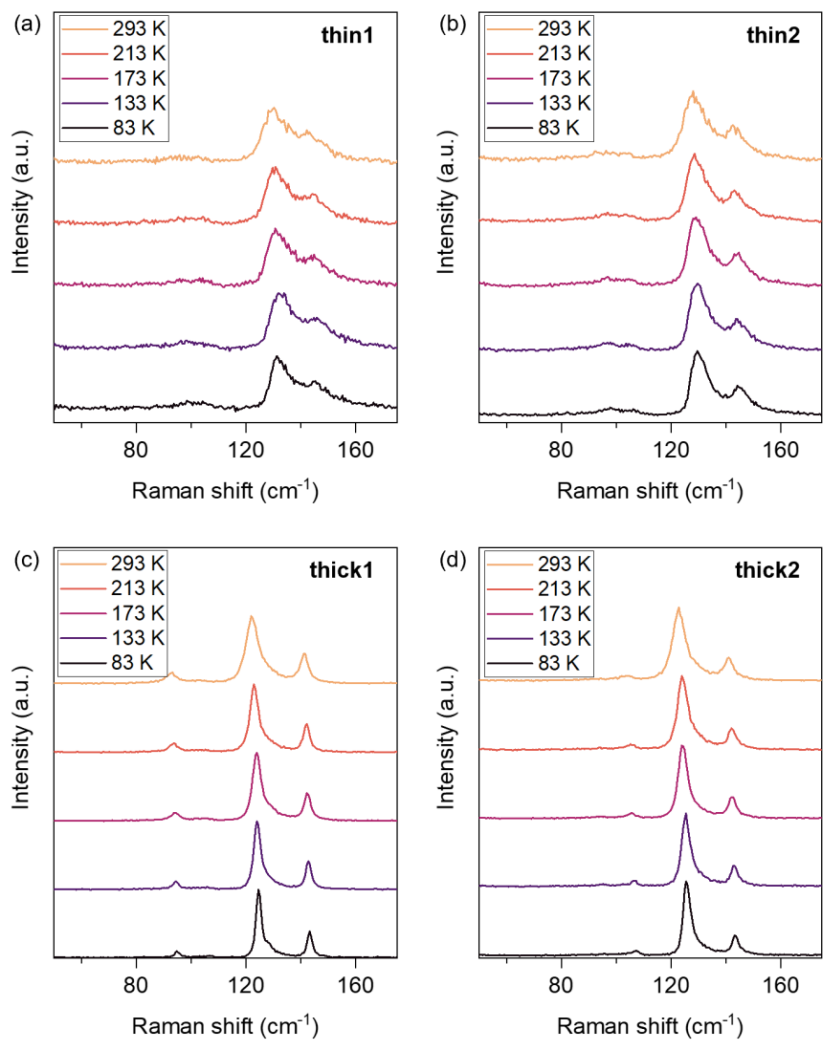

**Fig. S6**. Temperature-dependent Raman spectra for four Te flakes. The two labeled 'thin' are below the transition thickness, while the two labeled 'thick' are above the transition thickness based on AFM scans and their Raman features.

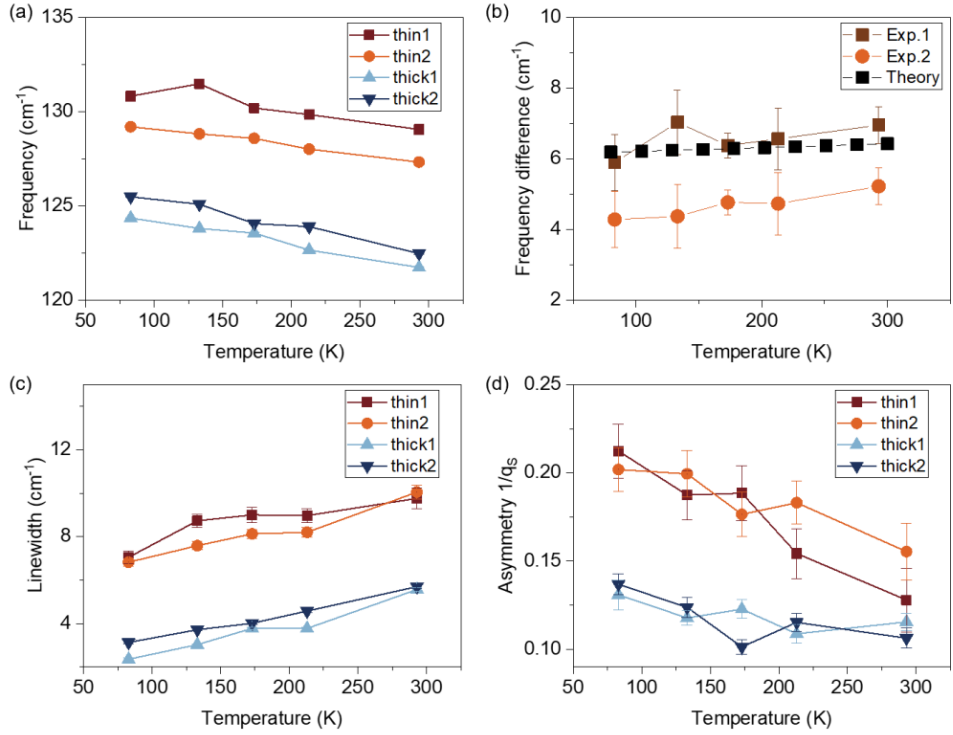

**Fig. S7**. Temperature dependence of the $A_1$ phonon mode for four Te flakes in **Figure S6**. (a) The experimental frequency as a function of temperature. (b) Frequency differences between thin and thick Te flakes as a function of temperature. Exp.1 (Exp.2): the difference between thin1 (thin2) and the average of thick1 and thick2. The error bar is the standard deviation of thick1 and thick2. Theory: theory-predicted frequency difference between small-polaron-dominated phonon and large-polaron-dominated phonon. The temperature dependence of the large-polaron-dominated phonon was taken from the fitting of the experimental thick Te. (c-d) The experimental linewidth and asymmetry as a function of temperature. The spectra are fitted with a BWF function. The error bar for the experimental data in (c-d) represents the fitting error.

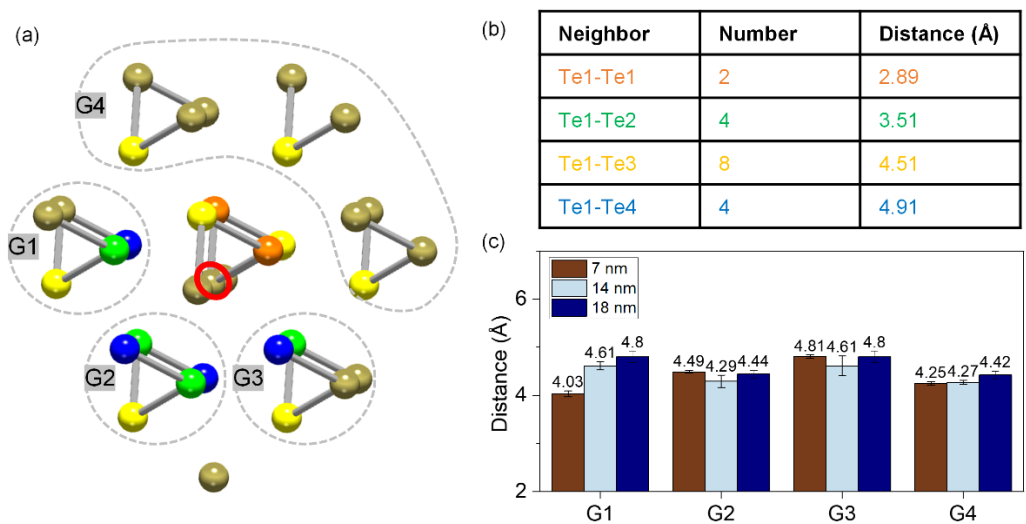

**Fig. S8**. Extended X-ray absorption fine structure (EXAFS) fitting model and results. (a) Illustration of the crystal structure and fitting model. The atoms colored in orange, green, yellow, and blue correspond to the atoms with a distance of 2.89, 3.51, 4.51, and 4.91 Å to the center atom circled in red. (b) The number of atoms within 5 Å to the center atom. A tellurium path of $R_{eff}$ = 2.89 Å, N = 2 is used for the nearest atoms in the same chain, and another tellurium path of $R_{eff}$ = 4.51 Å, N = 2 is used to account for the second nearest atom in the same chain. Fitting parameter $\alpha_i$ ($i$ = 1, 2) was given for each path to describe the expansion or contraction of the nearest and second-nearest Te-Te bonds. In terms of atoms in the three neighboring chains G1-G3 that may serve as scatters, tellurium paths of $R_{eff}$ = 3.51 Å (N = 4), 4.51 Å (N = 3), and 4.91 Å (N = 4) were assigned. For atoms in each chain, they were given the same $\alpha_i$ ($i$ = 3, 4, 5 for three chains) to evaluate the interchain distance. Group G4 contains three neighboring chains with the tellurium path of $R_{eff}$ = 4.51 Å (N = 3) and fitting parameter $\alpha_i$ ($i$ = 6). The model was fitted simultaneously to the data for three different thicknesses processed by using a *k*-weighting of 1 and 2 in the Fourier transform. (c) Fitted interchain distance between different helical chains (each chain is composed of a group of atoms as illustrated in (a)) labeled as G1 to G4 to the center helical chain.

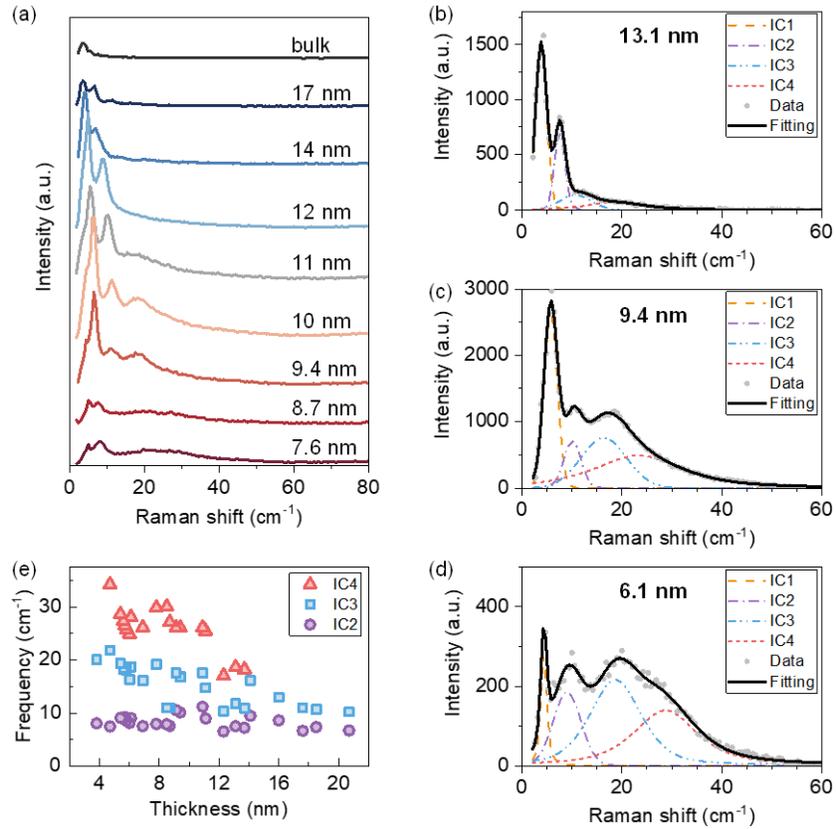

**Fig. S9.** Thickness-dependent interchain coupling. (a) Low-frequency Raman spectra of tellurene with different thicknesses. (b-d) Fitting of the low-frequency Raman modes for tellurene with a thickness of 13.1 nm, 9.4 nm, and 6.1 nm. (e) The Raman frequency of the interchain (IC) modes is summarized as a function of thickness. The interchain modes are labeled as IC1-IC4 based on their frequency for simplicity. The interchain mode IC1 has cut-off by the notch filter, thus, the frequency IC1 does not truly represent the phonon frequency.

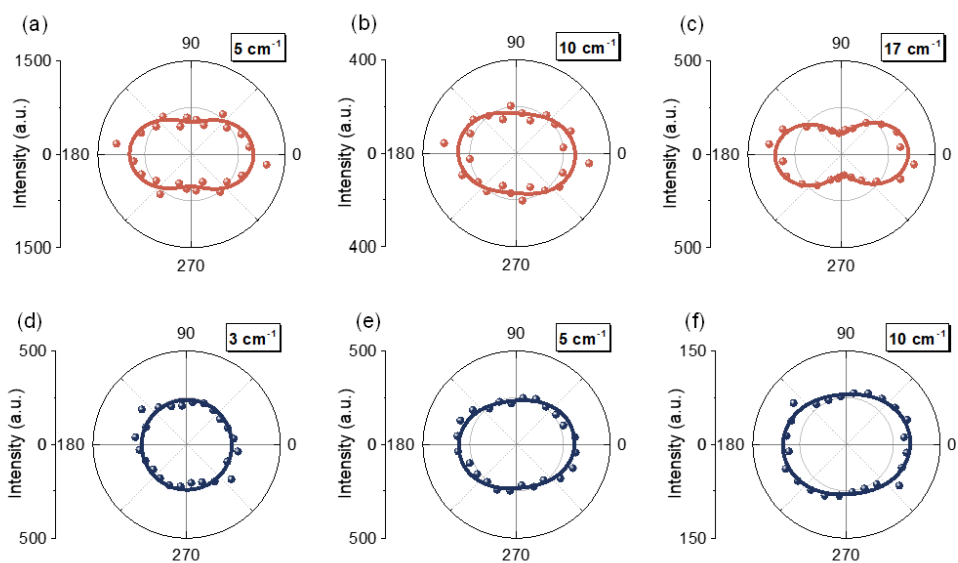

**Fig. S10**. Polar plots of the low-frequency Raman modes for tellurene of (a-c) 9.4 nm and (d-e) 16.9 nm. The sample was rotated in-plane and the scattering polarized parallel to the incident light was collected. The polarization angle is defined as the angle between the laser polarization and the [0001] direction.

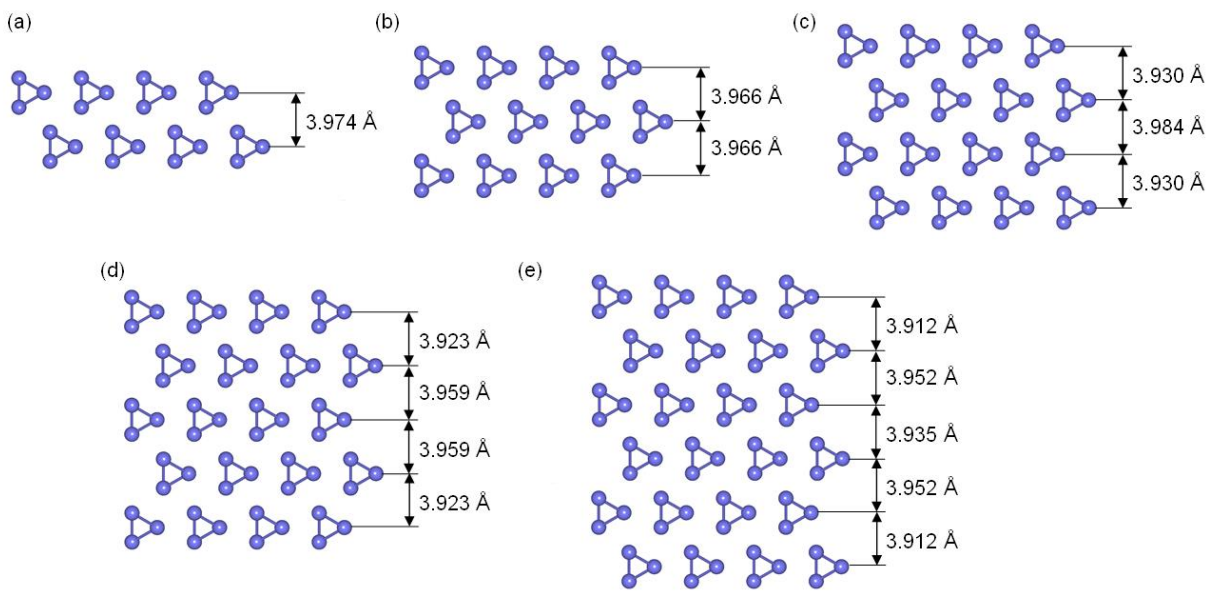

**Fig. S11**. The calculated vertical distance using first-principles calculations for tellurene with different numbers of layers. (a) 2L. (b) 3L. (c) 4L. (d) 5L. (e) 6L.

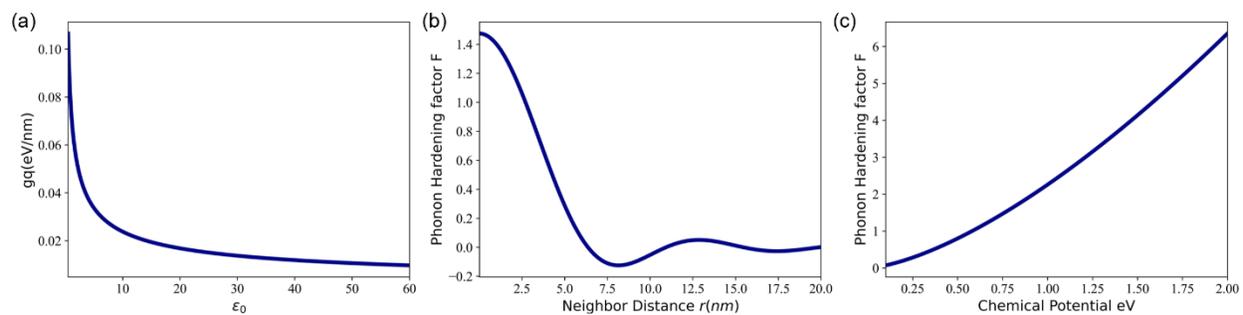

**Fig. S12.** (a) the static dielectric function dependent polar electron-phonon coupling coefficient after multiplying **q**. (b) the nearest neighbor distance-dependent phonon hardening factor. (c) the chemical potential dependent phonon hardening factor.

**Table S1**. Best fitting results of EXAFS. The fitting range was set as 1.85-5.00 for R-space and 3.0-6.5 for k-space. The $E_0$ is 6 for all fittings.

| | | | 7 nm | | 14 nm | | 18 nm | |
|---|---|---|---|---|---|---|---|---|
| Scattering pair | Coordination number | $R_{eff}$ | Debye-Waller factor | R | Debye-Waller factor | R | Debye-Waller factor | R |
| Te1-Te1 | 2 | 2.8923 | 0.00414 | 2.8712 | 0.0016 | 2.8439 | 0.00473 | 2.877 |
| Te1-Te3 | 2 | 4.5080 | 0.00598 | 4.2475 | 0.00229 | 4.4929 | 0.00683 | 4.7584 |
| Te1-Te2 | 1 | 3.5069 | 0.00508 | 3.1368 | 0.00196 | 3.5840 | 0.0058 | 3.7308 |
| Te1-Te3 | 1 | 4.5080 | 0.00598 | 4.0322 | 0.00229 | 4.6071 | 0.00683 | 4.7959 |
| Te1-Te4 | 1 | 4.9131 | 0.00614 | 4.3946 | 0.00235 | 5.0211 | 0.00702 | 5.2268 |
| Te1-Te2 | 2 | 3.5069 | 0.00508 | 3.4867 | 0.00196 | 3.3318 | 0.0058 | 3.4500 |
| Te1-Te3 | 1 | 4.5080 | 0.00598 | 4.4821 | 0.00229 | 4.2829 | 0.00683 | 4.4349 |
| Te1-Te4 | 2 | 4.9131 | 0.00614 | 4.8849 | 0.00235 | 4.6678 | 0.00702 | 4.8334 |
| Te1-Te2 | 1 | 3.5069 | 0.00508 | 3.7337 | 0.00196 | 3.5840 | 0.0058 | 3.7308 |
| Te1-Te3 | 1 | 4.5080 | 0.00598 | 4.7996 | 0.00229 | 4.6071 | 0.00683 | 4.7959 |
| Te1-Te4 | 1 | 4.9131 | 0.00614 | 5.2309 | 0.00235 | 5.0211 | 0.00702 | 5.2268 |
| Te1-Te3 | 3 | 4.5080 | 0.00598 | 4.2475 | 0.00229 | 4.2640 | 0.00683 | 4.4137 |